\documentclass[authoryear,preprint,review,11pt]{elsarticlemoda}
\usepackage{amsfonts}
\usepackage{amssymb}
\usepackage{amsmath}
\usepackage{setspace}
\usepackage{amsthm}
\usepackage{eucal}
\usepackage{bbm}
\usepackage{comment}
\usepackage{lineno}
\usepackage{rotating}
\usepackage{graphicx}
\usepackage{threeparttable}
\usepackage{adjustbox}
\usepackage[margin=1in]{geometry}
\usepackage{xcolor}
\usepackage{pgffor}
\usepackage{enumitem}
\usepackage{xr}

\newtheorem{theorem}{Theorem}
\newtheorem{lemma}[theorem]{Lemma}
\newtheorem{assumption}{Assumption}

\newtheorem{algorithm}{Algorithm}

\numberwithin{equation}{section}
\numberwithin{theorem}{section}
\numberwithin{condition}{section}
\numberwithin{definition}{section}
\theoremstyle{remark}
\newtheorem{remark}{Remark}
\journal{To appear in Econometric Reviews}

\begin{document}

\linespread{1.25}

\setlength{\abovedisplayskip}{3pt}
\setlength{\belowdisplayskip}{3pt}

\begin{frontmatter}



\title{Extending MinP Tests for Global and Multiple Testing}


\author{Zeng-Hua Lu}
\address{School of Economics, Adelaide University}
\ead{zen.lu@adelaide.edu.au}

\begin{abstract}
Empirical economic studies often involve multiple propositions or hypotheses, with researchers aiming to assess both the collective and individual evidence against these propositions or hypotheses. To rigorously assess this evidence, practitioners frequently employ tests with quadratic test statistics, such as $F$-tests and Wald tests, or tests based on minimum/maximum type test statistics. This paper introduces a combination test that merges these two classes of tests using the minimum $p$-value principle. We illustrate our proposed tests to re-evaluate two empirical applications reported in the literature. One concerns the evaluation of the effectiveness of a matching grant on charitable giving. The other examines the effectiveness of exercise.

\end{abstract}


\end{frontmatter}

\section{Introduction}

Empirical economic studies often involve multiple propositions or
hypotheses. In such studies, researchers are interested in evaluating the
collective and individual evidence against these propositions or hypotheses.
A prominent example in the recent literature is the evaluation of treatment
effects that are measured by several variables. Two popular classes of
statistical tests adopted by researchers are tests based on quadratic-form
test statistics (QF tests) or tests based on the minimum of $p$-values (or
the maximum of the corresponding test statistics). QF tests evaluate overall
treatment effects; they include the Wald, $F$- and Hotelling's $T^{2}$ tests
(see \cite{chagne09}, \cite{bendan16}, \cite{angjorkue18}, \cite{beaman21}
and \cite{chosutzim22} for examples). The minimum $p$-values (MinP) tests
evaluate the treatment effect relating to individual variable measurements,
and look to control the multiplicity of Type I errors (see \cite{anderson08}%
, \cite{leesha14}, \cite{gerhecetal14}, \cite{lishxu16} and \cite%
{chosutzim22} for examples).

Generally speaking, QF tests have good global power in detecting overall
treatment effects arising from the accumulation of many small individual
treatment effects. However, a rejection by a QF test does not indicate which
individual treatment effects are statistically significant, unless
subsequent testing via the closed testing procedure finds further evidence
(see, e.g., \cite{marpergar76}, \cite{roazwo11}, \cite{lu20cl} and \cite{goehemsol19}). In comparison, detection of any individual treatment effect
in MinP tests implies an overall treatment effect. This tends to provide a
global power advantage over QF tests when one or a few individual treatment
effects are present. MinP tests are popular in multiple testing; they can be
viewed as a special case of closed tests, but have the computational
advantage of sequential testing over general closed tests.

Naturally, the primary objective in the evaluation of treatment effects is
the presence of the overall treatment effect. However, the researcher
usually does not have a priori knowledge of the true distribution generating
the data in an empirical investigation. Since QF and MinP tests have
distinct global powers for detecting the presence of the overall treatment
effect, the researcher is likely to attempt both tests to examine empirical
evidence. For example, the researcher may attempt the other test if one test
does not provide the evidence they anticipates, or they may initially aim to
identify as many possible variables that relate to the treatment effect as
possible, but turn to QF tests after the initial attempt fails to identify
any variables. In fact, we argue that the researcher should attempt both
tests for a robustness check on how sensitive the empirical evidence is to
the distribution generating the data.

If both the tests are carried out, but only the results of one test are
reported, or a test is chosen based on sample information, then the error
probability of claimed findings may not be controlled at the desired level.
For example, if the researcher chooses to report the more significant
evidence on the overall treatment effect from the two tests attempted, then
the error probability of claiming positive findings when there are no
treatment effects (Type I error) becomes inflated beyond the level the
researcher intends to control. Over-reporting has drawn increasing concerns
in the literature. Recent papers by \cite{christensenmiguel18} and \cite%
{young19} provide excellent accounts of the issue. In his influential paper,
\cite{wh2000} draws attention to the dangerous practice of data snooping
with primary concern on testing predictive superiority when different
forecasting models are explored (see also \cite{hansenpr05}, \cite{romwol05}
and \cite{roshwo08}).

This paper proposes an adjusted $p$-value method which offers a tool for
robust and honest reporting. In fact, we show that the method can be viewed
as a combined test constructed through the minimum $p$ value principle,
which is a special case of the generalized mean of $p$-values for combining
tests (cf. \cite{vovkwang20} and \cite{vovkwang22}). We show that our
combined tests are admissible whenever both constituent tests are admissible.
Admissibility is an essential requirement for a testing procedure, as otherwise it would be dominated by another test (cf. \cite%
{lehrom05} and \cite{owen09}). Our result helps researchers construct tests
of the type we propose. Furthermore, we provide the necessary and sufficient
condition under which the combined test improves the global power of the
constituent tests. The combined test also retains the stepdown procedure of
MinP tests when simultaneously testing individual treatment effects, and
preserves the control of the familywise error rate (FWER) by the MinP test.

The $p$-values in the combined test can be easily computed based on the
Bonferroni correction; more precise adjustments may be computed through
resampling schemes (e.g., \cite{leesha14}, \cite{churom16} and \cite%
{bugcansha19}). Computational implementations are readily available in
popular packages such as R and Stata (see \cite{brhowe16} and \cite%
{claromwol20}), which can be easily extended to compute our adjusted $p$%
-values through the inclusion of an additional $p$-value associated with a
global test such as a QF test.

\textit{Related literature}. The idea of combining different tests can be
traced back to \cite{tippett31}, \cite{pearson33} and \cite{fisher36}. There
has been renewed interest in combining tests in the recent literature,
including approaches that use both global and maximum-type test statistics.
\cite{vovkwang20} and \cite{vovkwang22} studied a class of combination tests
based on the generalized mean of $p$-values and suggested the closed testing
procedure based on their combination tests for multiple testing. Our test is
a Tippett type test constructed based on the minimum of the $p$-values of
the constituent tests, and may be viewed as a special case of the
generalized mean of $p$-values. \cite{lu16} proposed an extended MaxT
(EMaxT) test by combining a sum test and a MaxT test for one-sided testing
where the sum test is chosen to direct to the `middle' of the constrained
parameter space under the global hypothesis. Our tests here can be viewed as
a generalization of the EMaxT test to a more general setting, by combining
p-values to allow for the adoption of a general global test, as well as
retaining a stepdown procedure of multiple testing. Both the EMaxT test and
the combined test proposed in this paper have a common feature of preserving
the rejection region shapes of constituent tests, hence somewhat inheriting
the respective strength of the constituent tests with regard to global
power. In fact, the properties studied in this paper also apply to the EMaxT
test. In high dimensional problems, combination tests of maximum type tests
and QF tests have drawn interests in the recent literature due to their
distinct global powers against sparse and dense alternatives. \cite%
{fanliayao15} proposed tests aiming at enhancing the global power of QF
tests in high-dimensional settings. Their tests combine a QF test and some
screened $t$-tests where the screened $t$-test components have an asymptotic
size $0$. \cite{kockpre23power} provide sufficient conditions for \cite%
{fanliayao15}'s test to achieve global power enhancement. \cite{kockpre23}
proposed a novel $p$-norm-based test that aims to achieve some global power
properties. \cite{hexuwupan21} and \cite{fejilixi20} derived approximated
distributions of constituent tests in high-dimensional settings, and studied
the properties of subsequent combination tests. However, these
high-dimensional combination tests are not devised for multiple testing, but
rather for global testing. There are also other combination tests proposed
in the recent literature. \cite{andrewsI16} studied a class of combination
tests for testing of weakly identified models by exploring global power
advantages of the two distinct tests in combination. \cite{helmeicha19}
constructed a combined test for multiple testing based on the marginal $p$%
-values conditional on a global test. Our paper contributes to this growing
body of literature by proposing a combined test that allows for global and
multiple testing simultaneously. In practice, the set of individual
hypotheses rejected by our combined test is the same as that of the
constituent MinP test whenever an individual hypothesis is rejected by our
combined test, while revealing a potentially much stronger signal in global
testing. This distinguishes our approach from the above recent combination tests that
aim at achieving some global power optimality. A further distinction of our tests is that they can serve as a tool for robust and honest reporting of global testing evidence and provide a decision guide when QF and MinP tests offer conflicting decisions.

When the number of hypotheses is large, the control of the FWER may be too
stringent. There have been other methods for controlling Type I errors
proposed in the literature, such as the $k$-FWER--the probability of
rejecting at least $k$ true null hypotheses, the false discovery proportion
(FDP)--the proportion of rejected null hypotheses that are actually false
with the default value 0 if there is no rejection, and the false discovery
rate (FDR)--the expected value of FDP (see, e.g., \cite{roshwo08}, \cite%
{rowo10} and \cite{harliu20}). Among these methods FDR originally proposed
by \cite{benhoc95} has been increasingly popular. It has been studied in
various econometric applications (see, e.g., \cite{gushen18}, \cite%
{harliuzhu16}, \cite{anderson08} and \cite{romwol05}). Recently, \cite%
{harliu20} argued that in the high-dimensional problems--where the number of
hypotheses is very large--taking into account both Type I and Type II errors
may be more appropriate in enhancing the power of positive discovery. They
propose several measures to balance the control between the two type errors.

The remainder of the paper is organized as follows. We motivate and
illustrate our combined test in the case of a multivariate normal location
model in Section \ref{s:ills}. We then present the combined test in a more
general setting in Section \ref{s:test} where some properties concerning the
combined test are established. In Section \ref{s:app} simulation studies are
provided to examine the performance of the combined test compared with other
tests. Section \ref{s:realapp} provides two real data applications.
Concluding remarks are made in Section \ref{s:concl}. Proofs are presented
in a supplementary file accompanying the paper.

\section{Illustrations}

\label{s:ills}

To fix ideas, we begin with an example of testing the multivariate normal
mean with a known covariance. Let $X=(X_{1},...,X_{k})^{\prime }\sim N(\mu
,\Sigma )$, where $\mu =(\mu _{1},...,\mu _{k})^{\prime }$ and $\Sigma $ has
the structure of the equicorrelation matrix $\{\rho _{ij}\}$, $i,j\in
K=\{1,\cdots ,k\}$, with $\rho _{ij}=\rho $, $-1<\rho <1$, when $i\neq j$
and $\rho _{ij}=1$, when $i=j$. The individual hypotheses are:%
\begin{equation*}
H_{i}:\mu _{i}=0,\quad vs\quad H_{i}^{\prime }:\mu _{i}\neq 0,\quad i\in K.
\end{equation*}%
Let $\Phi (\cdot )$ be the cumulative distribution function (CDF) of the
standard normal random variable and $F_{\chi _{k}^{2}}(\cdot )$ be the CDF
of the central chi-square random variable with $k$ degrees of freedom, and $%
F_{\chi _{k}^{2}}^{-1}(\cdot )$ be the inverse function of $F_{\chi
_{k}^{2}} $. Denote the individual $p$-value by $\hat{p}_{i}(X_{i})=2\Phi
(-\left\vert X_{i}\right\vert /\Sigma _{ii})$, where $\Sigma _{ii}$\ is the (%
$i,i$)-th element of $\Sigma $.

\subsection{A simple example}

\label{s:mexample} Consider the two-sided testing with $k=2$. With the
control of the FWER, the probability of rejecting at least one true $H_{i}$
(see Section \ref{s:test} for more details), MinP tests would reject $H_{i}$
if
\begin{equation*}
\hat{p}_{i}(X_{i})\leq c_{m}(\alpha ),
\end{equation*}%
where $\alpha \in (0,1)$ is the significance level, and $c_{m}(\alpha )$
satisfies
\begin{equation*}
\Pr_{\mu =0}[\hat{p}_{m}(X)=\min \{\hat{p}_{i}(X_{i}),i\in K\}\leq
c_{m}(\alpha )]=\alpha .
\end{equation*}%
Consider the global null hypothesis $H_{K}:\mu =0$ which is the intersection
of all individual null hypotheses $\cap _{i=1}^{k}H_{i}$. Rejection of any $%
H_{i}$ implies rejection of $H_{K}$. Therefore, MinP tests can be used to
jointly test $H_{K}$ against $H_{K}^{\prime }:\mu \neq 0$. They reject $%
H_{K} $ if any $\hat{p}_{i}(X_{i})\leq c_{m}(\alpha )$ or $\min \{\hat{p}%
_{i}(X_{i}),i\in K\}\leq c_{m}(\alpha )$.

If the researcher uses the Likelihood Ratio (LR) tests for testing $H_{K}$,
the test statistic $X^{\prime }\Sigma ^{-1}X$ follows the null distribution $%
\chi _{k}^{2}$ and $H_{K}$ would be rejected if
\begin{equation*}
\hat{p}_{g}(X)=1-F_{\chi _{k}^{2}}(X^{\prime }\Sigma ^{-1}X)\leq \alpha .
\end{equation*}

\begin{figure}[tbp]
\par
\begin{center}
\includegraphics[trim = 20mm 160mm 20mm 20mm,
clip,width=\linewidth]{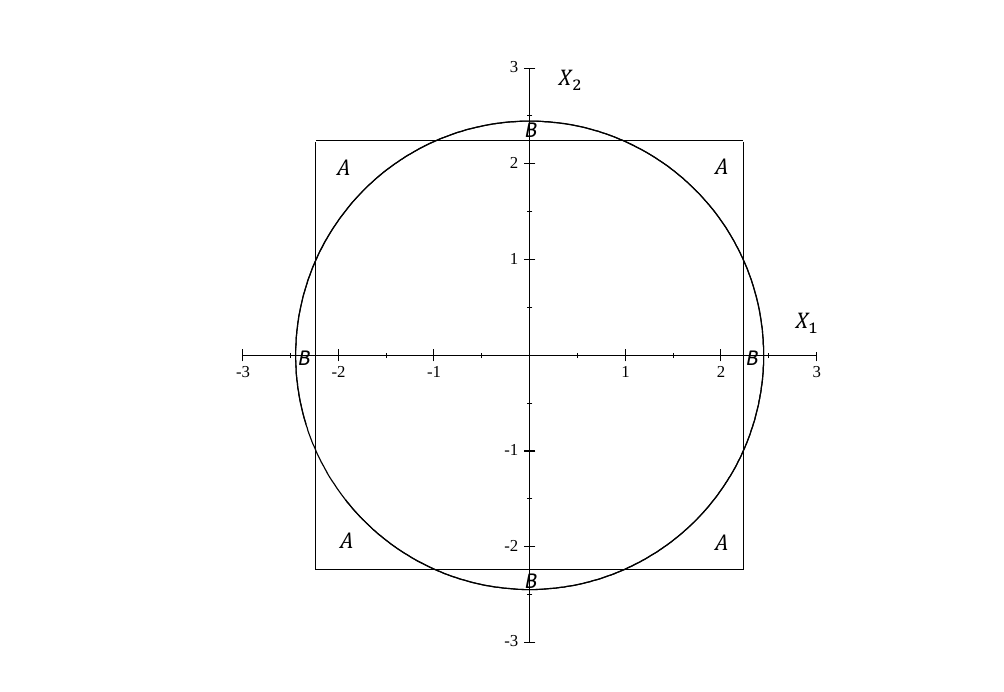}
\end{center}
\caption{A comparison of the rejection regions of LR (the area outside the
circle) and MinP tests (the area outside the square box) in the two-sided
testing of $\protect\mu =0$ when $\protect\rho =0$. }
\label{fh1}
\end{figure}

Figure \ref{fh1} shows the comparison of the rejection regions of LR tests
and MinP tests in the case of $k=2$ and $\rho =0$ with $\alpha =0.05$. The
rejection region of MinP tests is
\begin{equation*}
S_{m}(\alpha )=\{X:\hat{p}_{m}(X)\leq c_{m}(\alpha )\},
\end{equation*}%
where $c_{m}(\alpha )=1-(1-\alpha )^{1/2}=0.0253$. $S_{m}(\alpha )$
represents the area outside the square box. The rejection region of LR tests
is
\begin{equation*}
S_{g}(\alpha )=\{X:\hat{p}_{g}(X)\leq 0.05\},
\end{equation*}%
which is the area outside the circle. For $X\in A(\alpha )=S_{g}(\alpha
)\cap S_{m}^{c}(\alpha )$, where $S^{c}$ is the complement set of $S$, LR
tests reject $H_{K}$, but MinP tests do not reject $H_{K}$. For $X\in
B(\alpha )=S_{g}^{c}(\alpha )\cap S_{m}(\alpha )$, MinP tests reject $H_{K}$%
, but LR tests do not reject $H_{K}$. MinP tests are more likely to reject $%
H_{K}$ than LR tests when one of $\left\vert X_{i}\right\vert $, $i=1,2$,
dominates the other. However, as the boundary of the rejection region of LR
tests is defined by the circle $X_{1}^{2}+X_{2}^{2}=F_{\chi
_{2}^{2}}^{-1}(0.95)=5.99$, LR tests are more likely to reject $H_{K}$ than
MinP tests when none of $\left\vert X_{i}\right\vert $, $i=1,2$, dominates
the other. If one chooses LR tests or MinP tests based on the sample
information, it would inevitably lead to a data snooping problem. For
example, if one chooses LR tests or MinP tests depending on the outcome of
rejection, it would effectively lead to the enlarged rejection region as $%
S_{g}(\alpha )\cup B(\alpha )$ or $S_{m}(\alpha )\cup A(\alpha )$. An
enlarged rejection region implies an inflated size. For example, the
inflated size is about $0.07$ at $\alpha =0.05$ when $\rho =0.9$ or $-0.9$.

With regards to multiple testing of $H_{i}$, $i=1,2$, MinP tests reveal the
evidence on testing $H_{i}$ with the control of the FWER. However, a
rejection of $H_{K}$ by LR tests itself does not directly indicate which $%
H_{i}$ should be rejected. For example, when both $X_{i}$ take the value $%
1.8 $, $H_{K}$ is rejected by LR tests as $\hat{p}_{g}(X)\approx 0.02$, but
none of $H_{i}$ is rejected. One could proceed to the closure testing
procedure for multiple testing of $H_{i}$, where the rejection of an $H_{i}$
requires rejections of all $H_{K_{i}}$, $\{i\}\subseteq K_{i}\subseteq K$,
in a general case of $k\geq 2$. Thus, the computations required for carrying
out closed tests can be prohibitive when $k$ is large.

Another disadvantage of closed tests for multiple testing is their
conservatism in controlling the FWER. The rejection region of closed tests
in the case of $k=2$ is
\begin{equation*}
\{S_{g}(\alpha )\cap S_{1}(\alpha )\}\cup \{S_{g}(\alpha )\cap S_{2}(\alpha
)\}
\end{equation*}%
where $S_{i}(\alpha )=\{X:\hat{p}_{i}(X_{i})\leq \alpha \}$, $i=1,2$. This
rejection region is a strict subset of $S_{g}(\alpha )$; hence, the FWER of
closed tests based on $S_{g}(\alpha )$ is strictly less than $\alpha $. As $%
k $ increases, the FWER control becomes more conservative and consequently,
the capacity to detect false $H_{i}$ is reduced.

\subsection{A combination of two tests}

The motivations for proposing our tests are twofold. One is to maintain the
relative strength of the global power of the LR tests and MinP tests. The
other is to inherit the advantage of the stepdown procedure of MinP tests
for multiple testing. These two objectives can be achieved by combining LR
tests and MinP tests by taking the minimum of the $p$-values of the two
tests.

In the above $k=2$ case, it follows that
\begin{eqnarray*}
S_{g}(\alpha )\cup S_{m}(\alpha ) &=&\{X:\hat{p}_{g}(X)\leq \alpha \}\cup
\{X:\hat{p}_{m}(X)\leq \alpha \}, \\
&=&\{X:\hat{p}_{g}(X)\wedge \hat{p}_{m}(X)\leq \alpha \},
\end{eqnarray*}%
where $a\wedge b$ defines $\min (a,b)$, and we may use both the operations
interchangeably throughout the paper. To maintain the relative strength of
the global power of LR tests and MinP tests, one may preserve the shape of
the combined rejection region by adjusting it through an $\alpha ^{\prime
}<\alpha $ such that the combined rejection region has a probability of $%
\alpha $ under $H_{K}$, namely,
\begin{equation}
\Pr_{\mu =0}\{X\in S_{g}(\alpha ^{\prime })\cup S_{m}(\alpha ^{\prime
})\}=\Pr_{\mu =0}(\hat{p}_{g}(X)\wedge \hat{p}_{m}(X)\leq \alpha ^{\prime
})=\alpha .  \label{eqalphprm}
\end{equation}%
This implies that for an observed $X=x$, the adjusted $p$-value $\hat{p}%
_{l}^{adj}(x)$, $l=g,m,i$, can be computed as
\begin{equation*}
\hat{p}_{l}^{adj}(x)=\Pr_{\mu =0}(\hat{p}_{g}(X)\wedge \hat{p}_{m}(X)\leq
\hat{p}_{l}(x)).
\end{equation*}%
One may compute the null distribution
\begin{equation*}
F_{c}(x)=\Pr_{\mu =0}(\hat{p}_{g}(X)\wedge \hat{p}_{m}(X)\leq x)
\end{equation*}%
based on Monte Carlo simulation by randomly drawing $X$ from $N(0 ,\Sigma ) $
if $\Sigma$ is assumed known, or through a resampling scheme such as
permutation or bootstrap. One may opt to ease the computational burden by
computing the adjusted $p$-values based on, for example, the Bonferroni
inequality. More detailed discussions are presented in Section \ref{s:test}.

Let $\hat{p}_{c}^{adj}(X)=\hat{p}_{g}^{adj}(X)\wedge \hat{p}_{m}^{adj}(X)$.
The rejection region based on the combined test is
\begin{equation*}
S_{c}(\alpha )=\{X:\hat{p}_{c}^{adj}(X)\leq \alpha \},
\end{equation*}%
which is equivalent to
\begin{equation*}
S_{c}(\alpha )=\{X:X\in S_{g}(\alpha ^{\prime })\cup S_{m}(\alpha ^{\prime
})\},
\end{equation*}%
where $\alpha ^{\prime }<\alpha $ satisfies (\ref{eqalphprm}). The rejection
rule of the combined test is to reject $H_{K}$ if $\hat{p}_{c}^{adj}(x)\leq
\alpha $, not to reject it otherwise.

\emph{Stepdown procedure}. As for MinP tests in multiple testing, an
improved ability to reject more $H_{i}$, $i\in K$, is possible for the
combined test through a stepdown testing procedure. For example, for the
points
\begin{equation*}
X\in \{X:\hat{p}_{1}^{adj}(X_{1})\leq \alpha ,\hat{p}_{2}^{adj}(X_{2})>%
\alpha ,\hat{p}_{2}(X_{2})\leq \alpha \}
\end{equation*}%
in the bivariate example, the combined test rejects $H_{1}$, but not $H_{2}$%
. However, if we proceed in the same fashion as in the stepdown procedure of
MinP tests, one would then reject $H_{2}$ in the second step because $\hat{p}%
_{2}(X_{2})\leq \alpha $. Although the combined test has a disadvantage
compared with MinP tests in the first step of multiple testing, the combined
test would have the same outcomes as MinP tests in the stepdown procedure if
the null hypothesis with the smallest $\hat{p}_{i}(X_{i})$ is rejected by
the combined test in the first step of the stepdown procedure. Algorithm \ref%
{al1} toward the end of the next section provides a detailed description of
the algorithmic procedure for implementing our tests.

\section{The combined test}

\label{s:test}

\subsection{Setup}

We study the combined test in a more general setting. Suppose that the
sample $X^{(n)}$, where $n$ indicates sample size, is generated from the
unknown distribution $P\in \mathbf{P}$, where $\mathbf{P}$ defines a set of
probability distributions. Let $\hat{p}=\hat{p}(X^{(n)})$ be a $p$-value.
Let $G^{(n)}(u,P)$, $u\in \lbrack 0,1]$, be a sequence of CDFs of $\hat{p}$
under $P\in \mathbf{P}$. Denote the test function by
\begin{equation*}
\phi ^{(n)}=\phi ^{(n)}(X^{(n)})=\mathbf{1(}\hat{p}_{c}^{adj}\leq \alpha ),
\end{equation*}%
where $\mathbf{1(\cdot )}$ is the usual indicator function. Note that for
ease of presentation, we consider nonrandomized tests in this paper.

Let $H_{i}$, $i\in K=\{1,...,k\}$, $k\geq 2$, be the individual null
hypotheses and $H_{i}^{\prime }$ be the corresponding alternative
hypotheses. Let the set of distributions under $H_{i}$ be $\mathbf{P}%
_{i}\subset \mathbf{P}$. Let $K_{i}\subseteq K$ be a sub-index set, and $%
K_{\ast }\subseteq K$ be the set containing the indices of true $H_{i}$.
(The subscript $i$ in $K_{i}$ will be useful when we discuss the stepdown
procedure later on). Denote by $\mathbf{P}_{K}=\cap _{i\in K}\mathbf{P}%
_{i}\subset \mathbf{P}$ the set of null distributions corresponding to the
global null hypothesis $H_{K}$ and by $\mathbf{P}_{K}^{\prime }=\mathbf{%
P\smallsetminus P}_{K}$ the set of distributions corresponding to the
alternative hypothesis $H_{K}^{\prime }$. Assume $\mathbf{P}_{K}\subset
\mathbf{P}_{i}$, for all $i\in K$. That is, $\mathbf{P}_{K}$ is a strict
subset of $\mathbf{P}_{i}$, for all $i\in K$.

Before proceeding, we pause to present our general hypothesis setup along
with testing examples. Let $P=P_{\theta }$, $\theta \in \Omega $, where $%
\Omega $ denotes the parameter space. For the Gaussian location model
illustrated in Section~\ref{s:ills}, $\theta =(\mu ,\sigma ^{2})$, and $\mathbf{P%
}_{i}=\{P_{\theta }:\mu _{i}=0\}$.

For the regression model, $y_{t}=\beta ^{\prime }x_{t}+\varepsilon _{t}$,
where $\beta =(\beta _{1},\ldots ,\beta _{d})^{\prime }\in \mathbf{B}$ represents the vector of
regression coefficients and $\varepsilon _{t}$ is a random variable with
mean $0$. We have $\Omega =\mathbf{B}\times \mathbf{\Xi }$, where $\mathbf{%
\Xi }$ is a nonparametric component defining the unknown distribution of $%
\varepsilon _{t}$. The null distribution is $\mathbf{P}_{i}=\{P_{\theta
}:\beta _{i}=0\}$, for $i=1,\ldots ,k$, with $k\leq d$.

A test $\phi ^{(n)}$ of $H_{K}$ is referred to as the asymptotic pointwise
level-$\alpha $ test if
\begin{equation*}
\limsup_{n\rightarrow \infty }E_{P\in \mathbf{P}_{K}}(\phi ^{(n)})\leq
\alpha ,
\end{equation*}%
where $E_{P\in \mathbf{P}_{K}}(\cdot )$ is the expected value with respect
to $P\in \mathbf{P}_{K}$. (In this paper, we restrict our attention to
pointwise control.) In relation to multiple testing, the FWER is the
probability of rejecting any $H_{i}$, $i\in K_{\ast }$, under the true $P\in
\mathbf{P}_{K_{\ast }}$. That is,
\begin{equation*}
\text{FWER}=\Pr_{P\in \mathbf{P}_{K_{\ast }}}(\text{reject at least one }%
H_{i},i\in K_{\ast }).
\end{equation*}%
The asymptotic pointwise FWER control at the level $\alpha $ based on the
sample $X^{(n)}$ is achieved if
\begin{equation*}
\limsup_{n\rightarrow \infty }\text{FWER} \leq \alpha .
\end{equation*}%
It is worth noting that any distribution restricted by a possible
configuration of true and null hypotheses belongs to $\mathbf{P}_{K_{\ast }}$%
. The above definition of the FWER control is known as strong control of the
FWER.

Since the true null set $K_{\ast }$ is typically unknown, nor is the true $%
P\in \mathbf{P}_{K_{\ast }}$, assuming the true $P\in \mathbf{P}_{K}$ (which
is referred to as the weak control in the literature) does not guarantee the
control of the FWER (see, e.g., \cite{romwol05b}). However, in many
applications, the researcher may be able to assume the subset pivotality
condition of \cite{wesyou93}, which says that the true $P\in \mathbf{P}%
_{K_{\ast }}$ is not affected by whether $H_{i}$, $i\in K\mathbf{%
\smallsetminus }K_{\ast }$, is true or not. Hence, the FWER can be
controlled by assuming the true $P\in \mathbf{P}_{K}$. The subset pivotality
condition is not a necessary condition for controlling FWER. \cite{romwol05b}
showed a weaker sufficient condition for controlling FWER that is also
satisfied under the subset pivotality condition. We shall show later that
the proposed combined test controls the FWER as long as the MinP test being
combined controls the FWER.

Let the subscripts $c,g$, $m$, and $i$ indicate the combined, global, MinP
and individual tests, respectively. The $p$-value of the combined test is
\begin{equation*}
\hat{p}_{c}=\hat{p}_{g}\wedge \hat{p}_{m}.
\end{equation*}%
The adjusted $p$-value of the combined test for testing $H_{K}$ is defined
as
\begin{equation*}
\hat{p}_{c}^{adj}=G_{c}^{(n)}(\hat{p}_{c},P\in \mathbf{P}_{K}),
\end{equation*}%
or equivalently,
\begin{equation}
\hat{p}_{c}^{adj}=\hat{p}_{g}^{adj}\wedge \hat{p}_{m}^{adj}.  \label{eqpt1}
\end{equation}%
where
\begin{equation*}
\hat{p}_{l}^{adj}=G_{c}^{(n)}(\hat{p}_{l},P\in \mathbf{P}_{K}\mathbf{)}%
,\quad l=g,m.
\end{equation*}%
The limiting null distribution of $G_{c}^{(n)}(\cdot ,P\in \mathbf{P}_{K}%
\mathbf{)}$ is usually unknown for our combined tests. However, we can
compute $G_{c}^{(n)}(\cdot )$ based on resampling methods such as bootstrap,
permutation, and subsampling.

The researcher may opt to ease the computational burden by computing
conservative adjusted $p$-values based on the Bonferroni inequality as%
\begin{eqnarray*}
\hat{p}_{g}^{adj} &=&2\hat{p}_{g}\wedge 1, \\
\hat{p}_{m}^{adj} &=&2\hat{p}_{m}\wedge 1=2k\min (\hat{p}_{i},i\in K)\wedge
1,
\end{eqnarray*}%
or
\begin{eqnarray*}
\hat{p}_{g}^{adj} &=&(k+1)\hat{p}_{g}\wedge 1, \\
\hat{p}_{m}^{adj} &=&\frac{k+1}{k}\hat{p}_{m}\wedge 1=(k+1)\min (\hat{p}%
_{i},i\in K)\wedge 1,
\end{eqnarray*}%
then compute $\hat{p}_{c}^{adj}$ based on (\ref{eqpt1}). One may improve the
conservativeness of the Bonferroni inequality for computing $\hat{p}%
_{m}^{adj}$ in testing $H_{K}$ by using inequalities such as the {\v{S}}id{%
\'{a}}k inequality (\cite{sidak68}) or the Simes inequality (\cite{simes86}
and \cite{sarkar98}). However, these inequalities are not generally
applicable.

\subsection{Global testing}

\label{s:single}

The combined test rejects the global null hypothesis $H_{K}$ if $\hat{p}%
_{c}^{adj}\leq \alpha $; otherwise it accepts $H_{K}$. We first present the
size property of the combined test. We refer to the procedure in which $\hat{%
p}_{c}^{adj}$ is computed based on $G_{c}^{(n)}(\cdot )$ as $\phi
_{c,1}^{(n)}$ and to the procedure based on the Bonferroni inequality as $%
\phi _{c,2}^{(n)}$. The $\phi _{c,2}^{(n)}$ procedure may be conservative in
the sense that the test size may be strictly less than $\alpha $, but it is
computationally easy to implement. Note that to keep the presentation
concise $\phi _{c}^{(n)}$ will be used to represent both $\phi _{c,1}^{(n)}$
and $\phi _{c,2}^{(n)}$ when no confusion is likely to arise.

\begin{assumption}
\label{a0} With fixed $P\in \mathbf{P}_{K}$ and $l\in \{m,g,c\}$, \newline
(i) $G_{l}^{(n)}(u,P\mathbf{)}\rightarrow G_{l}(u,P)$, \newline
(ii) $G_{l}(u,P)$ is a continuous and strictly increasing function of $u\in
\lbrack 0,1]$.\newline
(iii) $G_{l}(u,P)\leq u$, for all $u\in (0,1)$.
\end{assumption}

\begin{remark}
Assumption \ref{a0} relies on the validity of the particular resampling method
employed to implement our test. There are many references that discuss resampling methods, including regularity conditions to ensure the
satisfaction of this assumption, for example, \cite{efrTib93}, \cite%
{polRomWof99}, and \cite[Chapter~ 15]{lehrom05}.
\end{remark}

\begin{remark}
When $\mathbf{P}_{K}$ is a singleton, that is, testing the simple null
hypothesis, $G_{l}(u,P\in \mathbf{P}_{K})$ is typically the uniform
distribution on $[0,1]$. In such a case, Assumption \ref{a0}(iii) is met
with equality. When $\mathbf{P}_{K}$ is composite, Assumption \ref{a0}(iii)
requires stochastic domination by the uniform random variable to ensure size
control. For example, suppose $H_{i}:\mu _{i}\leq 0$, $i\in K$, in the
multivariate normal example in Section \ref{s:ills}, the null distribution $%
\mathbf{P}_{K}$ is composite. The least favourable null distribution at $\mu
=0$ for testing $H_{K}$ is stochastically dominated by $U(0,1)$ representing
the null distribution of the $p$-value with the true ${\mu }\in \{{\mu :}\mu
_{i}\leq 0,i\in K\}$.
\end{remark}

The following lemma concerns the size property of the combined test.

\begin{lemma}
\label{th21} 
(i) If Assumption \ref{a0} holds for $l=c$, then $\limsup_{n\rightarrow
\infty }E_{P\in \mathbf{P}_{K}}\phi _{c,1}^{(n)}\leq \alpha $. (ii) If
Assumption \ref{a0} holds for $l=m,g$, then $\limsup_{n\rightarrow \infty
}E_{P\in \mathbf{P}_{K}}\phi _{c,2}^{(n)}\leq \alpha $, with the inequality
holding strictly if
\begin{equation}
\liminf_{n\rightarrow \infty }E_{P\in \mathbf{P}_{K}}\phi _{l}^{(n)}(1-\phi
_{l^{\prime }}^{(n)})>0  \label{eqlm1}
\end{equation}
where $l,l^{\prime }\in \{g,m\}$ and $l\neq l^{\prime }$.
\end{lemma}

\begin{remark}
Assumption \ref{a0}(i) required for $l=c$ is to ensure that the estimation
of $G_{c}^{(n)}(\hat{p}_{c},P\mathbf{)}$ for the test $\phi _{c,1}^{(n)}$ is
consistent, while (\ref{eqlm1}) implies that $\phi _{l}^{(n)}$, $l\in
\{g,m\} $, have distinct limiting rejection regions.
\end{remark}

\begin{remark}
When two tests are attempted, adjusting the $p$-value is important for
overall control of test size, as illustrated in the simple example in
Section \ref{s:mexample}. In some cases, an inflated size can be more
severe. For example, \cite{birovami09} proposed the sum test for the global
testing of $H_{K}$. They showed that their tests have a power advantage when
individual means are equal. For the simple case of $k=2$, their test has the
test statistic $\left\vert X_{1}+X_{2}\right\vert $ and rejects $H_{K}$ if
it is greater than or equal to $\sqrt{2(1+\rho )}c_{\Phi }(\alpha )$, where $%
c_{\Phi }(\alpha )$ is the $(1-\alpha /2)$th quantile of the standard normal
distribution. Suppose that the researcher attempts their test and MinP
tests, but only reports the smallest $p$-value of the two tests, then the
inflated size is about 0.095 when $\alpha =0.05$ and $\rho =-0.9$, which is
nearly double the nominal size.
\end{remark}

\begin{remark}
While the test $\phi _{c,1}^{(n)}$ may be able to achieve the size control
asymptotically exactly at the level $\alpha $ in some cases, the
conservativeness of the test $\phi _{c,2}^{(n)}$ is reflected by their size
control being asymptotically strictly less than $\alpha $.
\end{remark}

We now turn to study some global properties. Consider a set of local
alternatives
\begin{equation*}
\mathbf{P}_{n,K}^{\prime }=\{P_{n}:P_{n}\rightarrow P_{0}\in \mathbf{P}%
_{K}\}\subset \mathbf{P}_{K}^{\prime }.
\end{equation*}

A test $\phi _{l}^{(n)}$ is asymptotically $d$-admissible if for any other
test $\phi ^{(n)}$
\begin{equation}
\limsup_{n\rightarrow \infty }E_{P}(\phi _{l}^{(n)})\leq
\liminf_{n\rightarrow \infty }E_{P}(\phi ^{(n)})\text{,\qquad }\forall P\in
\{P_{n}:P_{n}\in \mathbf{P}_{n,K}^{\prime }\}  \label{eqad1}
\end{equation}%
and
\begin{equation}
\liminf_{n\rightarrow \infty }E_{P}(\phi _{l}^{(n)})\geq
\limsup_{n\rightarrow \infty }E_{P}(\phi ^{(n)}),\text{\qquad }\forall P\in
\mathbf{P}_{K}  \label{eqad2}
\end{equation}%
jointly imply
\begin{equation*}
\lim_{n\rightarrow \infty }E_{P}(\phi _{l}^{(n)})=\lim_{n\rightarrow \infty
}E_{P}(\phi ^{(n)})
\end{equation*}%
for all $P\in \mathbf{P}$. The $d$-admissibility implies that $\phi
_{l}^{(n)}$ can have better global power for some $P_{n}\in \mathbf{P}%
_{n,K}^{\prime } $ asymptotically compared to any other test that do not
have an asymptotically larger size.

\begin{theorem}
\label{adth} Under Assumptions \ref{a0} if both tests $\phi _{g}^{(n)}$ and $%
\phi _{m}^{(n)}$ are asymptotically $d$-admissible, then the combined test $%
\phi _{c}^{(n)}$ is asymptotically $d$-admissible.
\end{theorem}

Although Theorem \ref{adth} states that both the test $\phi _{c,1}^{(n)}$
and $\phi _{c,2}^{(n)}$\ are asymptotically $d$-admissible, the power of $%
\phi _{c,2}^{(n)}$ may be asymptotically uniformly improved by $\phi
_{c,1}^{(n)}$. This improvement typically occurs when $\phi _{c,2}^{(n)}$
has a test size strictly less than the nominal level $\alpha $. In
comparison to $d$-admissibility, $\alpha $-admissibility is defined as
follows: for any other level-$\alpha $ test $\phi ^{(n)}$, (\ref{eqad1})
implies (\ref{eqad2}) for all $P\in \{P_{n}:P_{n}\in \mathbf{P}%
_{n,K}^{\prime }\}$ (\cite{lehrom05}). In other words, the $\alpha $%
-admissibility of a test demands that there does not exist any other test
that has better power for at least some $P\in \mathbf{P}_{K}$ and no worse
power for all other $P\in \mathbf{P}_{K}$.

\begin{theorem}
\label{adthb} Under Assumptions \ref{a0} if both the two tests $\phi
_{g}^{(n)}$ and $\phi _{m}^{(n)}$ are asymptotically $d$-admissible and $%
\lim_{n\rightarrow \infty }E_{P\in \mathbf{P}_{K}}(\phi _{c}^{(n)})=\alpha $%
, then the combined test $\phi _{c}^{(n)}$ is asymptotically $\alpha $%
-admissible.
\end{theorem}

\begin{remark}
One may construct $\phi _{m}^{(n)}$ based on the Bonferroni inequality and
combine it through a Monte Carlo or resampling method such that the combined
test achieves an asymptotically exact size $\alpha $. By Theorem \ref{adthb}
the resulting combined test $\phi _{c,1}^{(n)}$ is asymptotically $\alpha $%
-admissible.
\end{remark}

The following theorem provides a necessary and sufficient condition under
which the combined test enhances the power of a constituent test. Let $%
\tilde{\phi}_{l}^{(n)}=\mathbf{1}(\hat{p}_{l}^{adj}\leq \alpha )$, $l\in
\{g,m\}$, be the test function based on the adjusted $p$-value instead of
the unadjusted $p$-value used in $\phi _{g}^{(n)}$ and $\phi _{m}^{(n)}$.

\begin{theorem}
\label{th27}
\begin{equation*}
\liminf_{n\rightarrow \infty }E_{P\in \mathbf{P}_{n,K}^{\prime }}(\phi
_{c}^{(n)})>\limsup_{n\rightarrow \infty }E_{P\in \mathbf{P}_{n,K}^{\prime
}}(\phi _{l}^{(n)}),\qquad l\in \{g,m\},
\end{equation*}%
holds if and only if%
\begin{equation}
\liminf_{n\rightarrow \infty }E_{P\in \mathbf{P}_{n,K}^{\prime }}\psi
^{(n)}>0,  \label{eqga2}
\end{equation}%
where
\begin{equation*}
\psi ^{(n)}=(\tilde{\phi}_{l^{\prime }}^{(n)}-\tilde{\phi}_{l}^{(n)})\mathbf{%
1}(\tilde{\phi}_{l^{\prime }}^{(n)}>\tilde{\phi}_{l}^{(n)})-(\phi _{l}^{(n)}-%
\tilde{\phi}_{l}^{(n)}),
\end{equation*}%
$l\neq l^{\prime }$, $l,l^{\prime }\in \{g,m\}$.
\end{theorem}

The condition (\ref{eqga2}) indicates that the power gain to $\tilde{\phi}%
_{l^{\prime }}^{(n)}$ from $\tilde{\phi}_{l}^{(n)}$ is asymptotically
sufficient to offset the power loss of $\phi _{l}^{(n)}$ caused by the
adjustment of its $p$-value. This condition reveals the sources of the power
gain and loss in relation to the global power improvement of the combined
test. We illustrate the condition (\ref{eqga2}) in the bivariate example in
Section \ref{s:mexample}. Let $X_{i}=\sqrt{n}\hat{\mu}_{i}$, $i=1,2$, and $%
\sqrt{n}(\hat{\mu}-\mu )\sim N(0,\Sigma )$. Let $\tilde{S}_{g}(\alpha )=\{X:%
\hat{p}_{g}^{adj}(X)\leq \alpha ,\mu =0\}$ and $\tilde{S}_{m}(\alpha )=\{X:%
\hat{p}_{m}^{adj}(X_{i})\leq \alpha ,\mu =0\}$. Let
\begin{equation*}
\tilde{A}(\alpha )=\tilde{S}_{g}(\alpha )\cap \tilde{S}_{m}^{c}(\alpha ),
\end{equation*}%
\begin{equation*}
\tilde{B}(\alpha )=\tilde{S}_{g}^{c}(\alpha )\cap \tilde{S}_{m}(\alpha ).
\end{equation*}%
Without loss of generality we let $l=m$ and $l^{\prime }=g$. Then
\begin{equation*}
(\tilde{\phi}_{l^{\prime }}^{(n)}-\tilde{\phi}_{l}^{(n)})\mathbf{1}(\tilde{%
\phi}_{l^{\prime }}^{(n)}>\tilde{\phi}_{l}^{(n)})=\tilde{B}(\alpha )
\end{equation*}%
while
\begin{equation*}
\phi _{l}^{(n)}-\tilde{\phi}_{l}^{(n)}=1\quad \Longleftrightarrow \quad
S_{m}(\alpha )-\tilde{S}_{m}(\alpha ).
\end{equation*}%
If we decompose $S_{m}(\alpha )-\tilde{S}_{m}(\alpha )$ into the two
exclusive regions
\begin{equation*}
\Delta _{1}=\{S_{m}(\alpha )-\tilde{S}_{m}(\alpha )\}\cap \tilde{S}%
_{g}(\alpha ),\quad \Delta _{2}=\{S_{m}(\alpha )-\tilde{S}_{m}(\alpha
)\}\cap \tilde{S}_{g}^{c}(\alpha ),
\end{equation*}%
it follows%
\begin{eqnarray*}
&&\tilde{B}(\alpha )-(S_{m}(\alpha )-\tilde{S}_{m}(\alpha )) \\
&=&(\tilde{B}(\alpha )-\Delta _{1})-\Delta _{2} \\
&=&\{X:\phi _{m}=0,\tilde{\phi}_{g}=1\}-\{X:\phi _{m}=1,\tilde{\phi}_{m}=0,%
\tilde{\phi}_{g}=0\}.
\end{eqnarray*}%
where $(\tilde{B}(\alpha )-\Delta _{1})$ and $\Delta _{2}$ are mutually
exclusive. The rejection region for $\phi _{c}^{(n)}$ is
\begin{equation*}
\tilde{S}_{g}(\alpha )\cup \tilde{S}_{m}(\alpha )=\tilde{S}_{m}(\alpha )-%
\tilde{S}_{g}(\alpha )\cup \tilde{S}_{m}^{c}(\alpha ).
\end{equation*}%
For the combined test $\phi _{c}^{(n)}$ to exhibit a better global power in
testing $H_{K}$ under $P\in \mathbf{P}_{n,K}^{\prime }$ which corresponds to
the local alternatives $\sqrt{n}\mu $, the following condition must hold
\begin{equation*}
\Pr_{P\in \mathbf{P}_{n,K}^{\prime }}(X\in \tilde{S}_{g}(\alpha )\cup \tilde{%
S}_{m}(\alpha ))-\Pr_{P\in \mathbf{P}_{n,K}^{\prime }}(X\in S_{m}(\alpha
))>0,
\end{equation*}%
which is equivalent to
\begin{equation}
\Pr_{P\in \mathbf{P}_{n,K}^{\prime }}\{X\in (\tilde{B}(\alpha )-\Delta
_{1})\}-\Pr_{P\in \mathbf{P}_{n,K}^{\prime }}(X\in \Delta _{2})>0.
\label{eqga3}
\end{equation}%
When either $\mu _{i}$, $i=1,2$, dominates the other, the first term in the
left hand side of (\ref{eqga3}) is likely dominate the second term.
Consequently, the combined test improves the global power of the MinP test.

Theorem \ref{th27} demonstrates that the combined test can improve the
global power of the constituent tests under the condition stated in (\ref%
{eqga2}). For a $P\in \mathbf{P}_{n,K}^{\prime }$, this improvement is more
probable for the constituent test with a lower global power compared to the
other test. This result is formally articulated in the next theorem, which
establishes that the combined test exhibits a more balanced global power,
with its global power is bounded between that of two constituent tests.

\begin{theorem}
\label{thblndp} For $P\in \mathbf{P}_{n,K}^{\prime }$ and $l\neq l^{\prime }$%
, $l,l^{\prime }\in \{g,m\}$ if
\begin{equation*}
\liminf_{n\rightarrow \infty }E_{P\in \mathbf{P}_{n,K}^{\prime }}\psi
^{(n)}\geq 0,\quad \text{and\quad }\limsup_{n\rightarrow \infty }E_{P\in
\mathbf{P}_{n,K}^{\prime }}\psi ^{(n)}\leq \liminf_{n\rightarrow \infty
}E_{P\in \mathbf{P}_{n,K}^{\prime }}(\phi _{l^{\prime }}^{(n)}-\phi
_{l}^{(n)}),
\end{equation*}%
then,
\begin{equation}
\liminf_{n\rightarrow \infty }E_{P\in \mathbf{P}_{n,K}^{\prime }}(\phi
_{c}^{(n)})\geq \limsup_{n\rightarrow \infty }E_{P\in \mathbf{P}%
_{n,K}^{\prime }}(\phi _{l}^{(n)}),  \label{eqthblndp1a}
\end{equation}%
\begin{equation}
\limsup_{n\rightarrow \infty }E_{P\in \mathbf{P}_{n,K}^{\prime }}(\phi
_{c}^{(n)})\leq \liminf_{n\rightarrow \infty }E_{P\in \mathbf{P}%
_{n,K}^{\prime }}(\phi _{l^{\prime }}^{(n)}).  \label{eqthblndp1b}
\end{equation}
\end{theorem}

\begin{remark}
In the simulation studies reported in Sections \ref{s:ills} and \ref{s:app}
we found that in most cases, the global power of the combined test is
bounded between the global powers of two constituent tests.
\end{remark}

\subsection{Multiple testing}

\label{s:sd}

The combined test may be viewed as an extended MinP test by adding another $%
p $-value for testing the global $H_{K}$ to the minimand set of MinP tests.
If an individual null hypothesis $H_{i}$, $i\in K$, is rejected after it has
been adjusted for the extended multiplicity, then the combined test proceeds
to the usual stepdown procedure of MinP tests for multiple testing. It is
well known that the stepdown procedure has better power in rejecting false
individual null hypotheses than the single-step procedure (c.f. \cite%
{romwol05b}).

Let
\begin{equation*}
\hat{p}_{(1)}\leq \hat{p}_{(2)}\cdots \leq \hat{p}_{(k)}
\end{equation*}%
denote the ordered $p$-values $\hat{p}_{i}$, $i\in K$, and let $H_{(1)}$, $%
H_{(2)}$, ..., $H_{(k)}$, be the corresponding null hypotheses. If $\hat{p}%
_{m}^{adj}\leq \alpha $, which is equivalent to $\min (\hat{p}%
_{i}^{adj},i\in K)\leq \alpha $, where $\hat{p}_{i}^{adj}=G_{c}^{(n)}(\hat{p}%
_{i},P\in \mathbf{P}_{K}\mathbf{)}$, then reject $H_{(1)}$ and the combined
test proceeds to the usual stepdown procedure of MinP tests of the remaining
$H_{(2)}$, ..., $H_{(k)}$. Let $G_{m,K_{i}}^{(n)}(\cdot ,P\in \mathbf{P}%
_{K_{i}}\mathbf{)}$, $K_{i}\subset K$, be the CDF of $\min (\hat{p}_{i},i\in
K_{i})$. Let the adjusted $p$-value associated with testing $H_{(i)}$ be%
\begin{equation*}
\hat{p}_{m,(i)}^{adj}=G_{m,K_{i}}^{(n)}(\hat{p}_{(i)},P\in \mathbf{P}_{K_{i}}%
\mathbf{)},
\end{equation*}%
where $K_{i}=\{(i),...,(k)\}$. Then, usual MinP tests may be implemented for
$(i)=(2),...,(k)$: reject $H_{(i)}$ if $\hat{p}_{m,(i)}^{adj}\leq \alpha $,
stop otherwise. One may compute the adjusted $p$-values based on the
Bonferroni inequality as%
\begin{equation*}
\hat{p}_{m,(i)}^{adj}=\left\vert K_{i}\right\vert \min (\hat{p}_{i},i\in
K_{i})\wedge 1,
\end{equation*}%
where $\left\vert \cdot \right\vert $ is the cardinality of $K_{i}$.

One would expect the combined test to inherit properties of multiple testing
from MinP tests to some degree. Compared with the stepdown procedure of MinP
tests, our combined test differs only in the first step of rejecting $%
H_{(1)} $; the combined test rejects $H_{(1)}$ if $\hat{p}%
_{(1)}^{adj}=G_{c}^{(n)}(\hat{p}_{(1)},P\in \mathbf{P}_{K})\leq \alpha $
whereas MinP tests reject $H_{(1)}$ if $\hat{p}_{m,(1)}^{adj}=G_{m,K}^{(n)}(%
\hat{p}_{(1)},P\in \mathbf{P}_{K})\leq \alpha $. Since $G_{m,K}^{(n)}(u,P\in
\mathbf{P}_{K})\leq G_{c}^{(n)}(u,P\in \mathbf{P}_{K})$ for every $u\in
\lbrack 0,1]$, it follows $\hat{p}_{m,(1)}^{adj}\leq \hat{p}_{(1)}^{adj}$.
Thus, $\hat{p}_{(1)}^{adj}\leq \alpha $ implies $\hat{p}_{m,(1)}^{adj}\leq
\alpha $; a rejection of $H_{(1)}$ by the combined test implies the
rejection by the MinP test. Once $H_{(1)}$ is rejected by the combined test
both tests share the same stepdown procedure, hence share the same testing
outcome in terms of testing the remaining $H_{i}$, $i\in K\smallsetminus
\{(1)\}$.

The FWER control requires that the probability of rejecting at least one
true $H_{i}$, $i\in K_{\ast }$, is bounded above by the designated level $%
\alpha $. Because the true null set $K_{\ast }$ is typically unknown to the
researcher, the control of the FWER is not guaranteed in the stepdown
procedure. The control of the FWER for all possible configurations of true
and null hypotheses is referred to as the strong control of the FWER. \cite%
{romwol05b} showed that MinP tests control the FWER under a monotonicity
condition, which is a weaker condition than the subset pivotality condition
of \cite{wesyou93}. The next theorem shows that the combined test controls
the FWER as long as the MinP test being combined controls the FWER.

\begin{theorem}
\label{th23} If the limit superior of the FWER of the MinP test being
combined is less than $\alpha $, then the limit superior of the FWER of the
combined test is also less than $\alpha $.
\end{theorem}

\begin{remark}
As $\hat{p}_{g}\wedge \hat{p}_{m}\leq \hat{p}_{m}$, it follows $\hat{p}%
_{m}^{adj}\geq \hat{p}_{m}$. Consequently, the ability to reject $H_{(1)}$
by the combined test may be compromised. Nevertheless, such a compromise may
result in a marked improvement in the global power of testing $H_{K}$ by $%
\phi _{m}^{(n)}$.
\end{remark}

\begin{remark}
Theorem \ref{th23} establishes the pointwise control of the FWER. The result
holds because the combined test and the MinP test differ only in the first
step of the stepdown procedure with respect to multiple testing. Moreover,
the combined test is always less or equally likely to reject a true
hypothesis $H_{i}$, $i\in K_{\ast }$, for each $P\in \mathbf{P}_{K}$. In
general, achieving uniform FWER control requires stronger conditions.
\end{remark}

\subsection{A general combined procedure}

To facilitate the application of our proposed combined test, this section
summarizes the procedure below. Let $T_{g}(X^{(n)})$ and $T_{i}(X^{(n)})$, $%
i\in K$, be the test statistics for the global testing of $H_{K}$ and
individual testing of $H_{i}$, $i\in K$, respectively. Without loss of
generality we assume rejecting $H_{K}$ and $H_{i}$ by the large values of $%
T_{l}(X^{(n)})$, $l=g,i$. We may write $T_{l}$ instead of $T_{l}(X^{(n)})$
for simplicity of notation.

\begin{algorithm}
\label{al1} \mbox{}

\begin{enumerate}
\item Compute the sample test statistics $T_{l}$, $l=g,i$.

\item Compute the null distribution of the $p$-value based on
resampling from the original sample, \label{alitem}

\begin{enumerate}[label=\ref{alitem}\alph*]

\item Generate under $H_{K}$ simulated samples, $X_{b}^{(n)\ast }$, for $%
b=1,...,B$, either through resampling methods such as bootstrap,
permutation, and subsampling.

\item Compute the test statistics $T_{l,b}^{\ast }$, for $l=g,i$, $b=1,...,B$%
.

\item Compute $p$-values $\hat{p}_{l}=B^{-1}\sum\nolimits_{b=1}^{B}\mathbf{1(%
}T_{l}\leq T_{l,b}^{\ast })$ and $p_{l,b}^{\ast
}=B^{-1}\sum\nolimits_{b^{\prime }=1}^{B}\mathbf{1(}T_{l,b}^{\ast }\leq
T_{l,b^{\prime }}^{\ast })$, for $l=g,i$, $b=1,...,B$.
\end{enumerate}

or use the limiting distribution of the test statistics.

\begin{enumerate}[label=\ref{alitem}\alph*$^{\prime }$]

\item Generate under $H_{K}$ simulated statistics, $T_{l,b}^{\ast }$, $l=g,i$%
, $b=1,...,B$, from their limiting null distributions.

\item Compute $p$-values $\hat{p}_{l}$ and $p_{l,b}^{\ast }$ for $l=g,i$, $%
b=1,...,B$, using their limiting null distributions.
\end{enumerate}

\item Compute $p_{c,b}^{\ast }=\min (p_{1,b}^{\ast },\ldots ,p_{k,b}^{\ast
},p_{g,b}^{\ast })$, for $b=1,...,B$.

\item Compute the adjusted $p$-values $\hat{p}_{l}^{adj}=B^{-1}\sum%
\nolimits_{b=1}^{B}\mathbf{1(}\hat{p}_{l}\geq p_{c,b}^{\ast })$ for $l=g,i$,
and $\hat{p}_{m}^{adj}=\min (\hat{p}_{1}^{adj},\ldots ,\hat{p}_{k}^{adj})$.

\item Global testing: if $\hat{p}_{c}^{adj}=\hat{p}_{g}^{adj}\wedge \hat{p}%
_{m}^{adj}>\alpha $, stop; otherwise, reject $H_{K}$ and continue to the
next step.

\item Multiple testing: order $H_{(1)}$, $H_{(2)}$, ..., $H_{(k)}$ according
to $\hat{p}_{(1)}\leq \hat{p}_{(2)}\cdots \leq \hat{p}_{(k)}$. If $\hat{p}%
_{m}^{adj}>\alpha $, stop; otherwise, reject $H_{(1)}$ and continue to the
existing stepdown multiple testing procedure of MinP tests for the remaining
hypotheses $H_{(i)}$ as follows.

\begin{itemize}
\item For $i=2,...,k$, compute $\hat{p}_{m,(i)}^{adj}=B^{-1}\sum_{b=1}^{B}%
\mathbf{1(}\hat{p}_{(i)}\geq p_{m,(i),b}^{\ast })$, where $p_{m,(i),b}^{\ast
}=\min (p_{(i),b}^{\ast },\ldots ,p_{(k),b}^{\ast })$.

\item If $\hat{p}_{m,(i)}^{adj}>\alpha $, stop; otherwise, reject $H_{(i)}$
and continue.
\end{itemize}
\end{enumerate}
\end{algorithm}

\begin{remark}
Resampling methods such as the bootstrap and subsampling allow the
researcher to more actually take into account the dependent structure among
test statistics (see, e.g., \cite{roshwo08} and \cite{roshwo08b}).
\end{remark}

\begin{remark}
Our algorithm only requires one block of bootstrap samples to compute the
adjusted $p$-values; in the stepdown procedure the same set of bootstrapped $%
p_{m,(i),b}^{\ast }$ can be used. In some other methods a second resampling
scheme may be required. For example, for permutation tests proposed in \cite%
{churom16} they proposed a second bootstrap to deal with possibly different
covariances in permuted samples. \cite{harliu20} used the second bootstrap
to address unknown mean in measuring Type II errors in their approach.
\end{remark}

\section{Monte Carlo studies}

\label{s:app}

This section provides Monte Carlo studies on the performance of the combined
test and compares it to the performance of other tests. We first illustrate
it in the normal location model. We assume a known covariance $\Sigma $, so
power functions of tests can be well approximated for comparison. We assume
a known covariance matrix $\Sigma$, so the power functions of the tests can
be well approximated for comparison. We then assume $\Sigma$ is unknown and
consider the bootstrap method, either parametric or nonparametric.

\begin{figure}[tbp]
\begin{center}
\includegraphics[trim = 20mm 66mm 20mm 10mm,
clip,width=\linewidth]{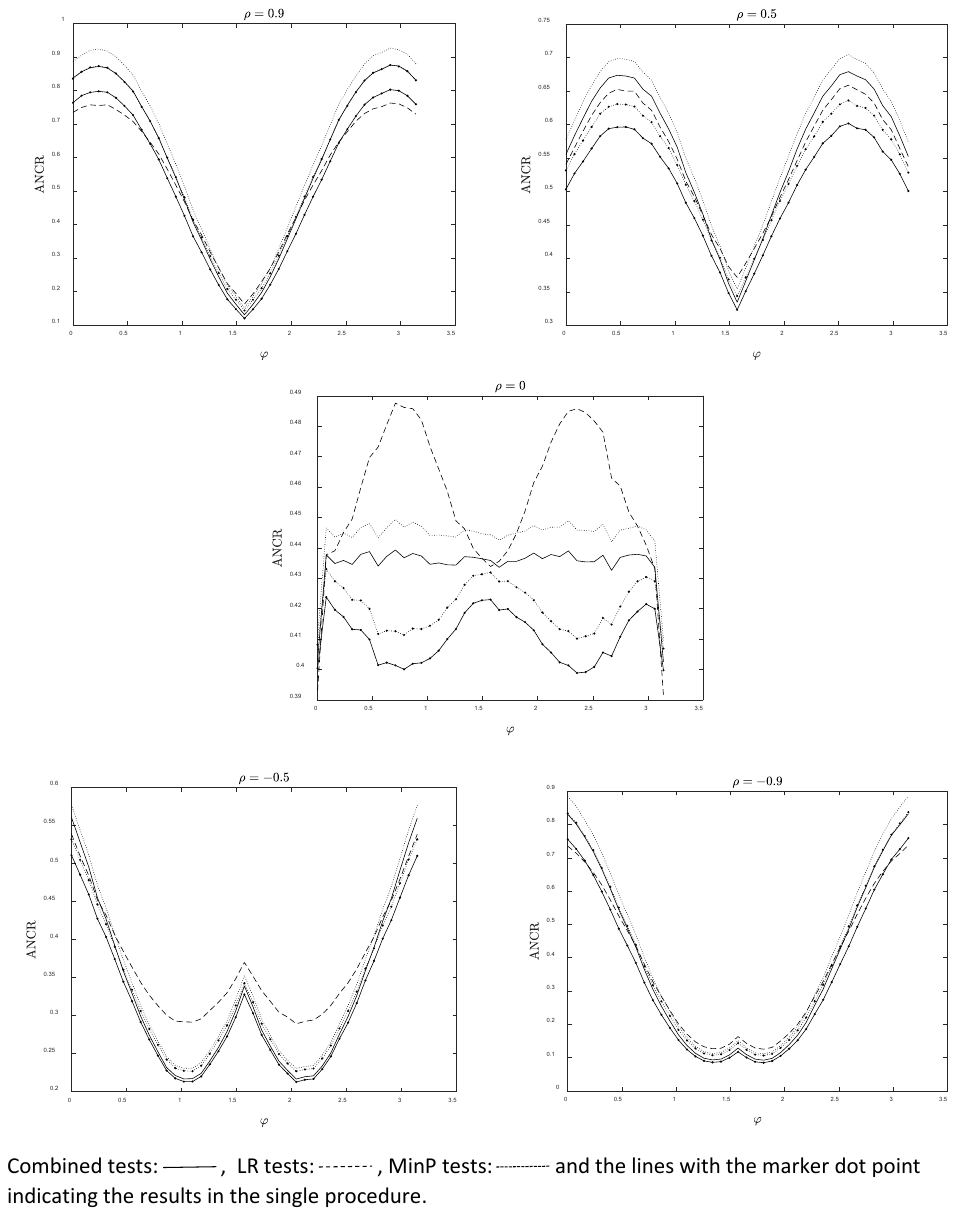}
\end{center}
\caption{A comparison of the average number of correctly rejected false $%
H_{i}$ by the combined, MinP and closed tests in the two-sided multiple
testing of $\protect\mu _{i}=0$, $i = 1,2$. }
\label{fh3}
\end{figure}

For the normal location models considered in Section \ref{s:ills}, assume
the known $\Sigma =\{\rho _{ij}\}$ with $\rho _{ij}=1$ when $i=j$ and $\rho
_{ij}=\rho $ when $i\neq j$. We approximate power functions of the combined
test and MinP tests based on $1,000,000$ and $100,000$ independent random
draws from $N(\mu ,\Sigma )$ for the cases of $k=2$ and $k\geqslant 2$,
respectively. The global power function of LR tests is computed as $\Pr
\{\chi _{k}^{2}(r^{2})>F_{\chi _{k}^{2}}^{-1}(0.95)\}$, where $\chi
_{k}^{2}(r^{2})$ is the chi-square random variable with $k$ degrees of
freedom and the non-centrality parameter $r^{2}$, and $F_{\chi
_{k}^{2}}(\cdot )$ denotes the CDF of the central chi-square distribution
with k degrees of freedom, and $F_{\chi _{k}^{2}}^{-1}(\cdot )$ is its
inverse function. The critical values of MinP tests and the combined test
can be approximated through random draws from $N(0,\Sigma )$. Figure \ref%
{fh2} presents the comparison of the global powers of testing $H_{K}$ for
LR, MinP and the combined test with $\alpha =0.05$ in the bivariate case.
For easy comparison we fix the power of LR tests at an $r$ value by taking $%
\mu _{1}=r\cos \varphi $, $r=2$ and
\begin{equation*}
\mu _{2}=\left\{
\begin{array}{c}
\rho \mu _{1}+\sqrt{(1-\rho ^{2})(r^{2}-\mu _{1}^{2})}\text{ \ \ \ if }\mu
_{1}\geq 0, \\
\rho \mu _{1}-\sqrt{(1-\rho ^{2})(r^{2}-\mu _{1}^{2})}\text{ \ \ \ if }\mu
_{1}<0,%
\end{array}%
\right.
\end{equation*}%
so that $\mu ^{\prime }\Sigma ^{-1}\mu =r^{2}$. The comparison shows that LR
tests have an overall global power advantage over MinP tests. However, MinP
tests can outperform LR tests when either $\left\vert \mu _{1}\right\vert $
or $\left\vert \mu _{2}\right\vert $ dominates the other. The combined test
somewhat inherits the respective strengths of LR and MinP tests with regard
to global power. In relation to multiple testing, the combined test can
outperform closed tests. However, MinP tests are more likely to reject $%
H_{i} $, $i\in K$, than the combined test. This is because $\hat{p}%
_{g}(x)\wedge \hat{p}_{m}(x)\leq \hat{p}_{m}(x)$ for every $x\in X$ with the
strict inequality holding for some $x\in X$. Figure \ref{fh3} presents a
comparison of the average number of correctly rejected (ANCR) false $H_{i}$
for closed, MinP and the combined test.

\begin{figure}[tbp]
\begin{center}
\includegraphics[trim = 20mm 195mm 20mm 9mm,
clip,width=\linewidth]{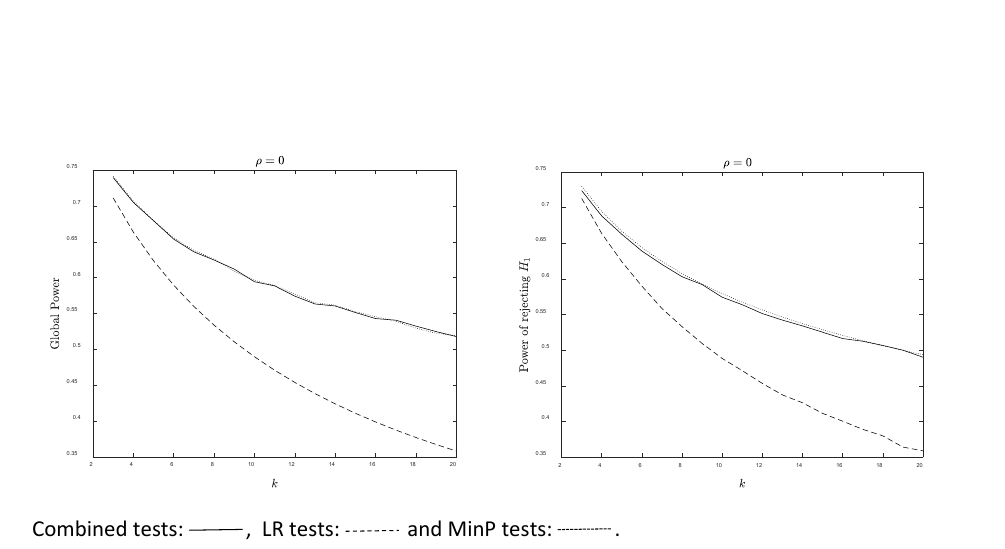}
\end{center}
\caption{A comparison of the global powers and the probabilities of
rejecting $H_{1}$ by the combined, LR/closed and MinP tests in two-sided
testing when $\Sigma =I$ and $\protect\mu =(3,0,...,0)^{\prime }$. }
\label{fh8}
\end{figure}

\begin{figure}[tbp]
\begin{center}
\includegraphics[trim = 20mm 195mm 20mm 9mm,
clip,width=\linewidth]{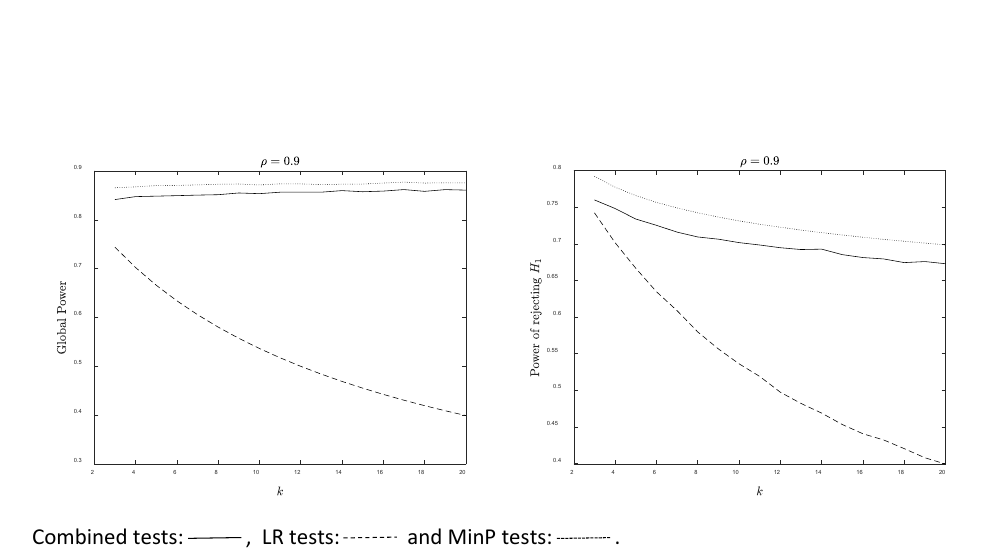}
\end{center}
\caption{A comparison of the global powers and the probabilities of
rejecting $H_{1}$ by the combined, LR/closed and MinP tests in two-sided
testing when $\protect\rho _{ij}=0.9$, $i\neq j$, $i,j\in K$, and $\protect%
\mu =(3,...,3)^{\prime }$. }
\label{fh9}
\end{figure}

To further illustrate that the combined test can share the power strength of
MinP tests to some extent, Figures \ref{fh8} and \ref{fh9} present
comparisons of the global power in testing $H_{K}$, as well as comparisons
of the probability of rejecting $H_{1}$ in multiple testing as the number of
hypotheses $k$ increases. The comparison in Figure \ref{fh8} is based on the
case of $\Sigma =I$ and $\mu =(3,0,...,0)^{\prime }$, while that in Figure %
\ref{fh9} is based on the case of $\rho _{ij}=0.9$, $i\neq j$, $i,j\in K$,
and $\mu =(3,...,3)^{\prime }$. The comparisons show that MinP tests have a
clear power advantage over LR tests in both global and multiple testing. The
advantage becomes increasingly apparent as $k$ increases. The combined test
in such cases shares some of the strength of MinP tests.

\begin{figure}[tbp]
\begin{center}
\includegraphics[trim = 20mm 73mm 20mm 10mm,
clip,width=\linewidth]{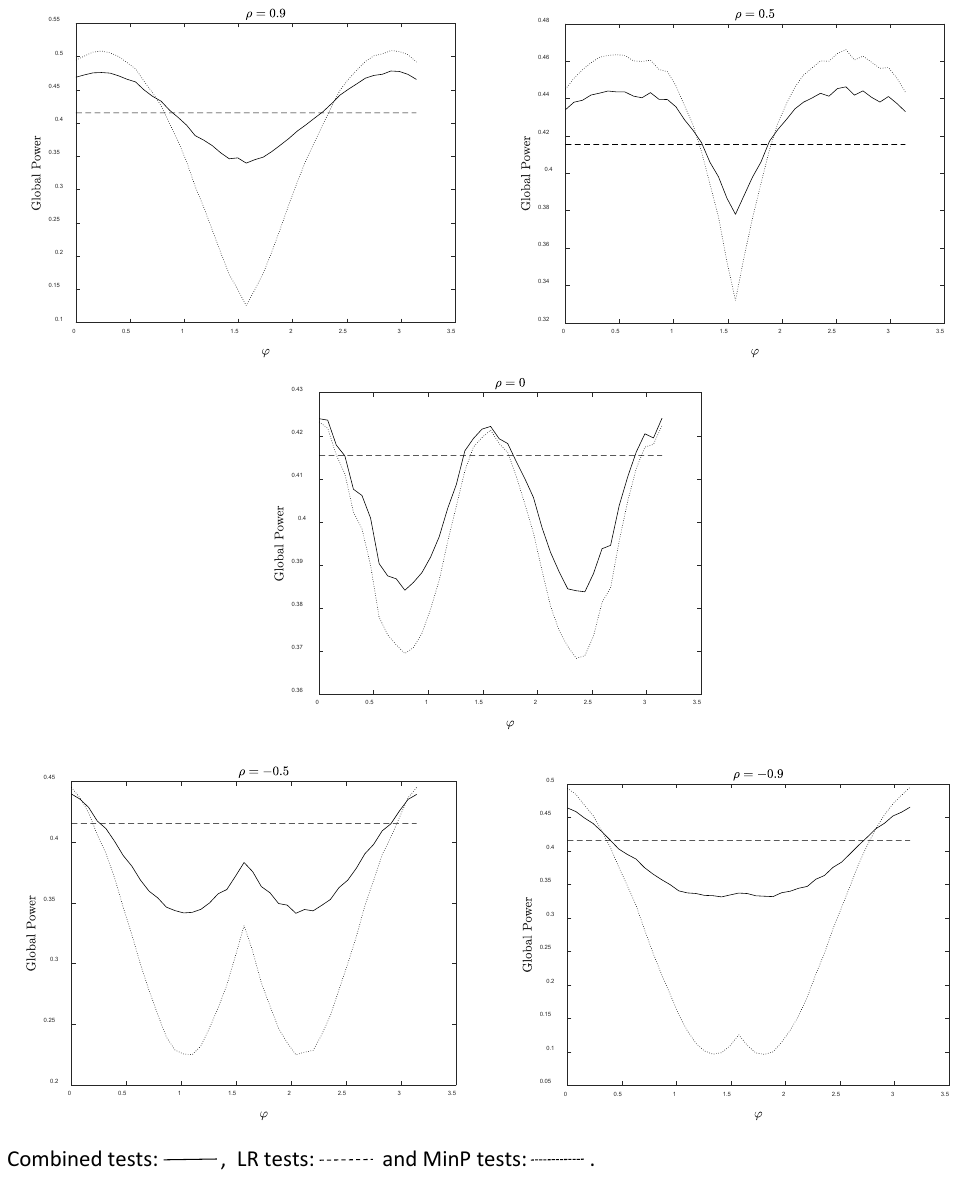}
\end{center}
\caption{A comparison of the global powers of the combined, LR and MinP
tests in the two-sided testing of $\protect\mu =0$. }
\label{fh2}
\end{figure}

Next we conduct simulation studies on tests of the multivariate mean on a
more realistic setting with an unknown covariance. Let $X^{(n)}=\{X_{t}%
\}_{t=1}^{n}$, where $X_{t}=(X_{t1},...,X_{tk})^{\prime }$ is an independent
$k$-dimensional random vector from the multivariate normal distribution with
the mean $\mu =(\mu _{1},...,\mu _{k})^{\prime }$ and covariance $\Sigma
_{X} $. Let $\bar{X}=(\bar{X}_{1},...,\bar{X}_{k})^{\prime }$ be the
studentized sample mean, and let $\hat{\Sigma}$ be the correlation matrix
corresponding to $\bar{X}$. By the multivariate central limit theorem it
follows that
\begin{equation*}
n^{\frac{1}{2}}\left( \bar{X}-\mu \right) \xrightarrow{d}N\left( 0,\Sigma
\right) ,
\end{equation*}%
where $\Sigma $ can be consistently estimated by $\hat{\Sigma}$.

We conducted simulation studies on two-sided testing of $\mu $ to compare
the performance of the $\phi _{c^{(1)}}^{(n)}$ test with that of other
tests. Hotelling's $T^{2}$ tests are adopted for testing the global null
hypothesis $H_{K}$, as well as for implementing the combined test and closed
tests. For an individual test of $H_{i}:\mu _{i}=0$ against $H_{i}^{\prime
}:\mu _{i}\neq 0$, $i\in K$, the test statistic is $T_{i}=\left\vert n^{%
\frac{1}{2}}\bar{X}_{i}\right\vert $. We consider the nonparametric and
parametric bootstrap approaches for computing $p$-values. In the
nonparametric bootstrap method, we generate the simulated samples $%
X_{b}^{(n)\ast }=\{X_{b,t}^{\ast }\}_{t=1}^{n}$ in Step $2a$ of Algorithm %
\ref{al1} by sampling uniformly at random with replacement from $X^{(n)}$,
and then recenter them at $\bar{X}$ by subtracting $\bar{X}$ from $%
X_{b,t}^{\ast }$ for all $t=1,\ldots ,n$. In the parametric bootstrap
method, we generate $\bar{X}_{b}^{\ast }$ in Step $2a^{\prime }$ of
Algorithm \ref{al1} from $N\left( 0,\hat{\Sigma}\right) $, then compute $p$%
-values $\hat{p}_{g,b}=1-F_{\chi _{k}^{2}}(n\bar{X}_{b}^{\ast \prime }\hat{%
\Sigma}^{-1}\bar{X}_{b}^{\ast })$ and $p_{i,b}^{\ast }=2\Phi (-\left\vert n^{%
\frac{1}{2}}\bar{X}_{i}\right\vert )$.

In our Monte Carlo simulation study the design of the correlation matrix $%
\Sigma _{X}=\left\{ \rho _{ij}\right\} $ has two structures. One takes the
form of the equicorrelation matrix, and the other takes $\rho
_{ij}=a^{\left\vert i-j\right\vert }$ with $a=0.5$ or $-0.5$. In the
nonparametric bootstrap approach $2,000$ bootstrap samples are used. In the
parametric bootstrap approach, we use $10,000$ random draws. The
significance level is set to $0.05$. The number of replications is set to $%
2,000$.

Table \ref{tb2} and Supp-Tables \ref{tbs1}--\ref{tb10} in the supplemental
file report the results of the simulation study. The results for the global
hypothesis testing are reported in terms of estimated sizes and powers under
the heading of $H_{K}$. The results for multiple hypothesis testing are
reported in terms of the estimated FWER and the estimated ANCR (average
number of correctly rejected false $H_{i}$). The notation $\mathbbm{1}_{k,m}$
represents the $k$-dimension column vector with the first $m$ elements being
$1$ and the remaining elements being $0$. We observe that Hotelling's $T^{2}$
tests outperform MinP tests in global power in many instances, but not in
all cases. In certain scenarios, the global power of MinP tests can be
considerably higher than that of Hotelling's $T^{2}$ tests. For example,
when the individual variables $X_{ti}$, $i \in K$, have equal means and are
equally positively correlated with $\rho_{ij} = 0.9$, the global power of
MinP tests is approximately 20\% and 30\% greater than that of Hotelling's $%
T^{2}$ tests for $k = 4$ and $k = 20$, respectively. This aligns with what
we observed in the previous case that assumed a known covariance $\Sigma $.
With regard to the ANCR false $H_{i}$ in multiple testing, the closure
procedure based on Hotelling's $T^{2}$ tests may outperform MinP tests in
some cases but not in others. The combined test appears to have its global
power bounded between the global powers of Hotelling's $T^{2}$ tests and
MinP tests in most cases. In some cases where $\mu _{i}$ for all $i$ are
equal--for example, when $\rho _{ij}=-0.25$ and $\mu _{1}=\ldots =\mu
_{4}=0.07$ in Table \ref{tb2}, the combined test considerably improves the
global power of MinP tests with little compromise on the multiple testing
power measured in terms of the ANCR false $H_{i}$. The results based on the
nonparametric bootstrap approach reported in Table \ref{tb2} and Supp-Table %
\ref{tbs1} are very close to the results based on the parametric bootstrap
approach reported in Supp-Tables \ref{tb3} and \ref{tb4} for the case $k=4$.

\begin{table*}[tbp]
\caption{The estimated sizes and powers (in percentages) in global testing,
and the estimated FWER (in percentages) and the ANCR (average number of
correctly rejected false $H_{i}$) in multiple testing procedures based on
bootstrap with $n=200$.}
\label{tb2}
\begin{center}
\tabcolsep=0.11cm
\par
\begin{tabular}{lccccccccc}
\hline
\multicolumn{1}{c}{$\mu$} & \multicolumn{3}{c}{Combined} &
\multicolumn{3}{c}{MinP} & \multicolumn{3}{c}{$T^{2}$} \\ \hline
& \multicolumn{1}{c}{$H_{K}$} & \multicolumn{1}{c}{FWER} &
\multicolumn{1}{c}{ANCR} & \multicolumn{1}{c}{$H_{K}$} & \multicolumn{1}{c}{
FWER} & \multicolumn{1}{c}{ANCR} & \multicolumn{1}{c}{$H_{K}$} &
\multicolumn{1}{c}{FWER} & \multicolumn{1}{c}{ANCR} \\ \hline
& \multicolumn{9}{c}{$\rho_{ij}=-0.25, i\neq j$} \\ \hline
$\mathbbm{1}_{4,0}$ & 4.80 & 4.45 & 0 & 4.70 & 4.70 & 0 & 4.75 & 2.20 & 0 \\
0.22$\mathbbm{1}_{4,1}$ & 83.20 & 1.40 & 0.72 & 74.60 & 1.45 & 0.73 & 88.85
& 3.20 & 0.68 \\
0.13$\mathbbm{1}_{4,2}$ & 80.55 & 2.15 & 0.45 & 48.55 & 2.20 & 0.46 & 90.35
& 4.45 & 0.49 \\
0.1$\mathbbm{1}_{4,3}$ & 88.30 & 1.10 & 0.40 & 42.45 & 1.15 & 0.41 & 95.20 &
2.35 & 0.50 \\
0.07$\mathbbm{1}_{4,4}$ & 76.45 & 0 & 0.26 & 26.45 & 0 & 0.26 & 88.80 & 0 &
0.31 \\ \hline
& \multicolumn{9}{c}{$\rho_{ij}=0, i\neq j$} \\ \hline
$\mathbbm{1}_{4,0}$ & 4.95 & 4.90 & 0 & 5.10 & 5.10 & 0 & 5.35 & 3.55 & 0 \\
0.25$\mathbbm{1}_{4,1}$ & 85.45 & 1.20 & 0.84 & 85.30 & 1.20 & 0.84 & 80.50
& 3.80 & 0.76 \\
0.2$\mathbbm{1}_{4,2}$ & 87.00 & 0.80 & 0.85 & 86.00 & 0.85 & 0.85 & 90.30 &
4.70 & 1.19 \\
0.15$\mathbbm{1}_{4,3}$ & 76.25 & 0.75 & 0.71 & 72.90 & 0.75 & 0.72 & 84.55
& 4.50 & 1.18 \\
0.14$\mathbbm{1}_{4,4}$ & 81.75 & 0 & 0.75 & 75.95 & 0 & 0.76 & 89.20 & 0 &
1.61 \\ \hline
& \multicolumn{9}{c}{$\rho_{ij}=0.5, i\neq j$} \\ \hline
$\mathbbm{1}_{4,0}$ & 5.50 & 5.05 & 0 & 5.35 & 5.35 & 0 & 4.95 & 2.60 & 0 \\
0.22$\mathbbm{1}_{4,1}$ & 84.10 & 1.35 & 0.73 & 74.90 & 1.40 & 0.74 & 89.30
& 3.20 & 0.78 \\
0.17$\mathbbm{1}_{4,2}$ & 78.65 & 1.00 & 0.65 & 66.65 & 1.15 & 0.66 & 84.90
& 3.05 & 0.98 \\
0.17$\mathbbm{1}_{4,3}$ & 81.45 & 0.70 & 0.72 & 74.35 & 0.70 & 0.74 & 86.10
& 5.25 & 1.50 \\
0.17$\mathbbm{1}_{4,4}$ & 78.45 & 0 & 0.78 & 79.45 & 0 & 0.79 & 67.40 & 0 &
1.98 \\ \hline
& \multicolumn{9}{c}{$\rho_{ij}=0.9, i\neq j$} \\ \hline
$\mathbbm{1}_{4,0}$ & 6.00 & 4.50 & 0 & 5.35 & 5.35 & 0 & 4.75 & 1.20 & 0 \\
0.11$\mathbbm{1}_{4,1}$ & 87.80 & 2.00 & 0.23 & 28.35 & 2.40 & 0.26 & 93.20
& 2.00 & 0.33 \\
0.09$\mathbbm{1}_{4,2}$ & 85.25 & 1.05 & 0.20 & 24.30 & 1.40 & 0.23 & 91.45
& 2.70 & 0.33 \\
0.1$\mathbbm{1}_{4,3}$ & 86.05 & 1.15 & 0.27 & 31.80 & 1.30 & 0.31 & 91.90 &
4.85 & 0.45 \\
0.2$\mathbbm{1}_{4,4}$ & 79.25 & 0 & 0.79 & 82.05 & 0 & 0.82 & 62.05 & 0 &
2.19 \\ \hline
& \multicolumn{9}{c}{$\rho_{ij}=0.5^{\left\vert i-j\right\vert }$} \\ \hline
$\mathbbm{1}_{4,0}$ & 5.35 & 5.00 & 0 & 5.35 & 5.35 & 0 & 5.00 & 2.90 & 0 \\
0.24$\mathbbm{1}_{4,1}$ & 85.95 & 1.30 & 0.80 & 81.90 & 1.45 & 0.80 & 88.10
& 2.40 & 0.79 \\
0.2$\mathbbm{1}_{4,2}$ & 82.20 & 1.25 & 0.78 & 80.40 & 1.30 & 0.79 & 83.00 &
4.45 & 1.24 \\
0.18$\mathbbm{1}_{4,3}$ & 82.95 & 0.65 & 0.80 & 80.95 & 0.70 & 0.80 & 82.30
& 4.35 & 1.64 \\
0.17$\mathbbm{1}_{4,4}$ & 82.15 & 0 & 0.82 & 82.90 & 0 & 0.83 & 78.65 & 0 &
2.15 \\ \hline
& \multicolumn{9}{c}{$\rho_{ij}=(-0.5)^{\left\vert i-j\right\vert }$} \\
\hline
$\mathbbm{1}_{4,0}$ & 5.65 & 5.60 & 0 & 5.80 & 5.80 & 0 & 5.40 & 2.75 & 0 \\
0.24$\mathbbm{1}_{4,1}$ & 88.00 & 1.10 & 0.82 & 84.50 & 1.10 & 0.83 & 90.60
& 2.95 & 0.81 \\
0.13$\mathbbm{1}_{4,2}$ & 73.40 & 1.65 & 0.49 & 51.95 & 1.80 & 0.50 & 85.85
& 2.90 & 0.53 \\
0.11$\mathbbm{1}_{4,3}$ & 83.20 & 0.55 & 0.46 & 48.40 & 0.55 & 0.48 & 92.60
& 1.90 & 0.62 \\
0.09$\mathbbm{1}_{4,4}$ & 78.55 & 0 & 0.38 & 39.05 & 0 & 0.39 & 90.25 & 0 &
0.58 \\ \hline
\end{tabular}%
\end{center}
\end{table*}

\section{Empirical applications}

\label{s:realapp} We demonstrate our proposed methods based on two empirical
studies reported in the literature. Denote by $Y_{q,r,t}$ the outcome
variable for the $r$th outcome in the $q$th group for the $t$th participant.
Denote the sample mean of the $q$th group by $\bar{Y}_{q}=n_{q}^{-1}%
\sum_{t=1}^{n_{q}}Y_{q,t}$, where $Y_{q,t}$ is a column vector of outcomes
for the $t$th participant in the $q$th group and $n_{q}$ is the sample size
in the $q$th group. Let $\hat{\Sigma}_{q}$ be the sample covariance and $%
\upsilon _{q}$ be the population mean of $Y_{q,t}$. The $r$th element of $%
\upsilon _{q}$ is denoted by $\upsilon _{q,r}$.

\subsection{Effectiveness of matching donations}

The first empirical study evaluates the effectiveness of a matching grant on
charitable giving studied by \cite{lishxu16}. The data and Matlab code used
to replicate their reported results are available at the following address:
https://github.com/seidelj/mht. The sample includes 50,083 individuals who
had given to the organization at least once since 1991. Each individual was
independently assigned with a two-thirds probability to the treatment group
and a one-third probability to the control group. More details can be found
in \cite{lishxu16} and \cite{karlanlist07}. In this application $Y_{q,r,t}$
represents the amount of individual donations in dollars for the $t$th
individual in the $q$th treatment group with $q=1,2,3$ corresponding to the
matching amounts of one, two and three dollars, respectively, and $q=0 $
corresponding to the control group with no matching contribution.

The first case we study is the evaluation of the effectiveness of a matching
grant on charitable giving based on the four outcomes of interest: the
response rate, dollars given not including the matching amount, dollars
given including the matching amount, and the change in the amount given (not
including the matching amount). In this case, $r\in K=\{1,...4\}$ and we
redefine $q=T$ when $q\in \{1,2,3\}$. We are interested in testing the
multiple hypotheses:
\begin{equation*}
H_{i}:\upsilon _{T,r}=\upsilon _{0,r}\quad \mathnormal{vs}\quad
H_{i}^{\prime }:\upsilon _{T,r}\neq \upsilon _{0,r},\quad i=r\in K.
\end{equation*}%
We estimate the covariance matrix of $\bar{X}=\bar{Y}_{q}-\bar{Y}_{0}$ by $%
\hat{\Sigma}_{X}=n_{T}^{-1}\hat{\Sigma}_{T}+n_{0}^{-1}\hat{\Sigma}_{0}$.

The second case we study is the evaluation of the effectiveness of three
matching amounts from matching donors on individual donations, with $Y_{q,t}$
being a scalar measuring dollars given not including the matching amount as
the outcome of interest. The multiple hypotheses of interest are
\begin{equation*}
H_{i}:\upsilon _{q}=\upsilon _{0}\quad \mathnormal{vs}\quad H_{i}^{\prime
}:\upsilon _{q}\neq \upsilon _{0},\quad i=q\in \{1,2,3\}.
\end{equation*}%
Redefine $\bar{X}=(\bar{Y}_{1}-\bar{Y}_{0},\bar{Y}_{2}-\bar{Y}_{0},\bar{Y}%
_{3}-\bar{Y}_{0})^{\prime }$. The covariance matrix of $\bar{X}$ is $\hat{%
\Sigma}_{X}=(\hat{\Sigma}_{X,ij})$, $i,j\in K$, with
\begin{equation*}
\hat{\Sigma}_{X,ij}=\left\{
\begin{array}{cc}
n_{q}^{-1}\hat{\sigma}_{q}^{2}+n_{0}^{-1}\hat{\sigma}_{0}^{2} & \text{if }i=j
\\
n_{0}^{-1}\hat{\sigma}_{0}^{2} & \text{if }i\neq j%
\end{array}%
\right. .
\end{equation*}%
In both the cases, \cite{lishxu16} suggested MaxT tests with the component
test statistic, $1-\hat{p}_{i}$, where $\hat{p}_{i}$ is the $p$-value of $t$
tests for testing $H_{i}$ based on $\bar{X}_{i}$. This MaxT test is
equivalent to MinP test. We adopt the Hotelling's $T^{2}$ type test
statistic $T_{g}=\bar{X}^{\prime }\hat{\Sigma}_{X}^{-1}\bar{X}$ for global
testing of
\begin{equation*}
H_{K}:\cap _{i=1}^{K}H_{i}\quad \mathnormal{vs}\quad H_{K}^{\prime }:\text{%
at least one }H_{i}\text{ is not true.}
\end{equation*}

Following \cite{lishxu16} we employ the bootstrap method to implement our
combined tests, as well as MinP tests. With their Matlab program to generate
the same $3000$ bootstrap samples, Table \ref{tbe2} presents the differences
in the sample averages $\bar{X}$ and $p$-values under different tests.

The headings `FS' and `SD' in the last two columns of the table represent
the adjusted $p$-values of our combined tests in the first step and the
step-down procedure, respectively. Note that only the smallest adjusted $p$%
-values of our combined tests and MinP tests are different because once they
proceed to the following steps of $p$-value adjustments they are adjusted in
the same way. Looking at the other results presented in the table, the
differences in the sample averages are identical to what is reported in
Table 1 of \cite{lishxu16}.

The unadjusted $p$-values reported under the heading `$T^{2}$/$t$' and the $%
p $-values for MinP tests are almost the same as theirs with one exception:
the unadjusted $p$-values for the outcomes `Response rate' and `Dollars
given including match' are $0$ in our results and $0.0003$ in theirs.

There is also a slight difference in the adjusted $p$-values of MinP tests
between our and theirs in the case of multiple treatment evaluation; the
comparison between the control group and the treatment group of `2:1'
matching is slightly different, with ours being $0.1300$ and theirs being $%
0.1297$. These differences are due to their coding error in using the Matlab
function `find', which gives inaccurate results when the observed sample $p$%
-value is $0$, or when there are tied bootstrapped $p$-values equal to the
observed sample $p$-value.

In relation to the effectiveness of a matching grant on charitable giving
measured by the four outcomes of interest, both $T^{2}$ and MinP tests show
sufficiently strong evidence. MinP tests further identify the outcome
variables `the response rate' and `dollars given not including the matching
amount' are significant contributors, while closed tests based on $T^{2}$
tests fail to identify any significant contributor. In this case, our
combined tests show the same evidence as MinP tests after accounting for the
multiplicity of test introduced by including $T^{2}$ tests. In relation to
the effectiveness of the three different matching amounts, all three
tests--namely, $T^{2}$, MinP tests, and our combined tests--show no
significant evidence.

\begin{table*}[tbp]
\caption{Differences in group averages ($\bar{X}$) and $p$-values for the
matching donations data. }
\label{tbe2}
\begin{center}
\tabcolsep=0.11cm
\par
\begin{tabular}{lcccccc}
\hline
&  &  &  &  & \multicolumn{2}{c}{Combined} \\ \hline
& \multicolumn{1}{c}{$\bar{X}$} & \multicolumn{1}{c}{$T^{2}$/$t$} &
\multicolumn{1}{c}{MinP} & \multicolumn{1}{c}{Closed} & \multicolumn{1}{c}{FS
} & \multicolumn{1}{c}{SD} \\ \hline
& \multicolumn{6}{c}{Multiple outcomes} \\ \hline
$T^{2}$ &  & 0 &  &  & 0 &  \\
Response rate & 0.0042 & 0 & 0 & 0.1457 & 0 & 0 \\
Dollars given not including match & 0.1536 & 0.0500 & 0.0963 & 0.2663 &
0.1547 & 0.0963 \\
Dollars given including match & 2.0876 & 0 & 0 & 0.1223 & 0 & 0 \\
Amount change & 6.3306 & 0.7200 & 0.7200 & 0.7520 & 0.9840 & 0.7200 \\ \hline
& \multicolumn{6}{c}{Multiple Treatments} \\ \hline
$T^{2}$ &  & 0.2260 &  &  & 0.5053 &  \\
Control versus 1:1 & 0.1234 & 0.2627 & 0.2627 & 0.3530 & 0.5590 & 0.2627 \\
Control versus 2:1 & 0.2129 & 0.0477 & 0.1300 & 0.2260 & 0.1367 & 0.1367 \\
Control versus 3:1 & 0.1245 & 0.2060 & 0.3533 & 0.3530 & 0.4713 & 0.3533 \\
\hline
\end{tabular}%
\end{center}
\end{table*}

\subsection{Effectiveness of exercise}

The second empirical study evaluates the effectiveness of exercise, as
studied by \cite{churom16} using the data reported in \cite{chagne09}. The
data are available as part of the supplementary materials to \cite{chagne09}
on the Econometrica website and contain seven biometric measures of the
participants. The measures are indicated in Table \ref{tbe1}. The
participants are randomly divided into three groups: the control group (G1),
the first treatment group who were paid \$25 to attend the gym once a week
(G2), and the second treatment group who were paid an additional \$100 to
attend the gym eight more times in the following four weeks (G3). G1 had 39
participants, G2 had 56 participants (after excluding one who had incomplete
observations on some of the variables) and G3 had 60 participants. See \cite%
{chagne09} for more details.

In this study $Y_{q,r,t}$ defines the change from the initial measurement
level to the final measurement level taken after 20 weeks for the $q$th
group, the $t$th participant and the $i$th measure with $q\in \{G1,G2,G3\}$,
$r\in K=\{1,...,7\}$ and $n_{G1}=39$, $n_{G2}=56$, $n_{G3}=60$. \cite%
{churom16} studied permutation tests of multiple hypotheses
\begin{equation*}
H_{i}:\upsilon _{q_{1},r}=\upsilon _{q_{2},r}\quad \mathnormal{vs}\quad
H_{i}^{\prime }:\upsilon _{q_{1},r}\neq \mu _{q_{2},r},\quad i=r\in K,
\end{equation*}%
where $q_{1}$, $q_{2}\in \{G1,G2,G3\}$ and $q_{1}\neq q_{2}$, for each $r\in
K$ across seven biometric measures. They showed that the covariance in the
permuted samples is not asymptotically equivalent to the covariance in the
observed sample unless two sample sizes are equal, i.e., $%
n_{q_{1}}=n_{q_{2}} $ or two population covariances are equal, $\Sigma
_{q_{1}}=\Sigma _{q_{2}}$. They suggested the closed testing procedure for
multiple hypothesis testing based on either their modified Hotelling's $%
T^{2} $ test or their MinP test, although they did not use the term `MinP'
as we do. We illustrate the combined test of their two tests for
constructing the closed testing procedure and compare it to their tests. We
also illustrate our combined tests in the step-down testing procedure under
the equal population condition.

The modified Hotelling's $T^{2}$ test has the test statistic%
\begin{equation*}
T_{g}=n_{q_{1}}\bar{X}_{q_{1}q_{2}}^{\prime }\hat{\Sigma}_{q_{1}q_{2}}^{-1}%
\bar{X}_{q_{1}q_{2}},
\end{equation*}%
where $\bar{X}_{q_{1}q_{2}}=\bar{Y}_{q_{1}}-\bar{Y}_{q_{2}}$ and $\hat{\Sigma%
}_{q_{1}q_{2}}=\hat{\Sigma}_{q_{1}}+\frac{n_{q_{1}}}{n_{q_{2}}}\hat{\Sigma}%
_{q_{2}}$, for testing
\begin{equation*}
H_{K}:\upsilon _{q_{1}}=\upsilon _{q_{2}}\quad \mathnormal{vs}\quad
H_{K}^{\prime }:\upsilon _{q_{1}}\neq \upsilon _{q_{2}}.
\end{equation*}%
The individual test of $H_{i}$ uses the test statistic%
\begin{equation*}
n_{q_{1}}^{1/2}\frac{\left\vert \bar{X}_{q_{1}q_{2},i}\right\vert }{\sqrt{%
\hat{\Sigma}_{q_{1}q_{2}}^{(ii)}}},\quad i\in K,
\end{equation*}%
where $\bar{X}_{q_{1}q_{2},i}$ is the $i$th element of $\bar{X}_{q_{1}q_{2}}$
and $\hat{\Sigma}_{q_{1}q_{2}}^{(ii)}$ is the $(i,i)$th element of $\hat{%
\Sigma}_{q_{1}q_{2}}$. For conducting closed tests one needs to test the
local intersection hypothesis $H_{J}$ for all $J\subseteq K$ using only $%
Y_{q,r,t}$, $r\in J$, to construct the test statistics.

Following \cite{churom16}, we generated the permuted samples. For each
permuted sample we bootstrapped $\hat{p}_{g}$ and $\hat{p}_{i}$ by following
Algorithm 2.1 of \cite{churom16}. The adjusted sample $p$-values, $\hat{p}%
_{g}^{adj}$ and $\hat{p}_{i}^{adj}$ were then computed as the proportions of
the values in the permuted sample sequence $\{\hat{p}_{c}(X^{d}),d\in D\}$
that were less than or equal to the observed $\hat{p}_{g}$ and $\hat{p}_{i}$%
, respectively. The number of random permuted samples was set to $10,000$
(with $9999$ permuted samples generated plus the original sample). The
number of nonparametric bootstrap samples was set to $3000$.

We compared the results of the combined test with those of Chung and
Romano's modified Hotelling's $T^{2}$ test and the MinP test, as well as the
closed tests based on the modified Hotelling's $T^{2}$ test and the MinP
test. Table \ref{tbe1} presents the difference in the sample averages $\bar{X%
}_{q_{1}q_{2}}$ (column 2), the associated standard errors (s.e.) $%
n_{q_{1}}^{-1/2}\sqrt{\hat{\Sigma}_{q_{1}q_{2}}^{(ii)}}$ (column 3) and $p$%
-values. The $p$-values of the single-step MinP test are reported in column
4: they were computed as the proportion of $\{\min (\hat{p}_{i}(X^{d}),i\in
K),d\in D\}$ that was less than or equal to the original sample $\hat{p}_{i}$%
. The $p$-values of the modified Hotelling's $T^{2}$ test are also reported
in column 4. The $p$-value of the closed tests for testing $H_{i}$ is
reported as the largest $p$-value of those obtained in testing $H_{J}$ for
all $J\subseteq K$ that involve $H_{i}$. Columns 5 and 6 report the $p$%
-values of the closed tests based on the modified Hotelling's $T^{2}$ test
and the MinP test, respectively. The adjusted $p$-values in the first step
of the combined test are reported in column 7. The adjusted $p$-values of
the combined test in the stepdown procedure are reported in column 8, where
the adjustment was computed as the proportion of $\{\min (\hat{p}%
_{i}(X^{d}),i\in K_{i}),d\in D\}$ that was less than or equal to the
smallest original sample $\hat{p}_{(i)}$ for each $(i)=(2),...,(k)$. We note
that the $p$-values of the modified Hotelling's $T^{2}$ test and the MinP
test reported here are slightly different from those reported in \cite%
{churom16}. These differences may be attributed to the fact that \cite%
{churom16} appeared to have used 57 participants in the G2 group, while we
used 56 participants. The difference may also be due to different random
numbers in generating random permuted and bootstrap samples.

Although group comparisons revealed different dominating effects, we
observed that effects on some biometric measures dominated in all three
group-wise comparisons. Therefore, it is not surprising that the MinP test
tended to have a smaller $p$-value than the modified Hotelling's $T^{2}$
test in testing the global hypothesis $H_{K}$. However, we cannot determine
if the stronger evidence presented by the MinP test was simply due to the
sampling variation. The combined test, which combines the modified
Hotelling's $T^{2}$ test and the MinP test, acts as a robust and honest
check.

When comparing the control group with the first treatment group, the MinP
test found insufficient evidence to suggest a difference between the two
groups, specifically in body fat and pulse rate measures. In contrast, the
modified Hotelling's $T^{2}$ test found insufficient evidence to suggest any
difference between the two groups. The combined test revealed that the
differences in body fat and pulse rate measures dominated those of the other
measures, although they were not statistically significant. In comparing the
control group with the second treatment group, the combined test, the MinP
test and the modified Hotelling's $T^{2}$ test found significant evidence to
suggest a difference between the two groups. Furthermore, all of the
multiple testing procedures--namely the combined test, the MinP test and the
closed procedures based on the modified Hotelling's $T^{2}$ test--indicated
that the effect on the body fat measure was significant. When comparing the
first treatment group with the second treatment group, the MinP test and the
modified Hotelling's $T^{2}$ test again yielded conflicting evidence in
rejecting $H_{K}$; the modified Hotelling's $T^{2}$ test found no
significant evidence, whereas the MinP test found significant evidence.
However, unlike in the case of comparing the control group with the first
treatment group, the combined test confirmed the findings of the MinP test,
along with the closed procedure based on the MinP test, in both the global
and multiple hypothesis tests.

\begin{table*}[tbp]
\caption{The differences in group averages, standard errors and $p$-values
for the exercise data. }
\label{tbe1}
\begin{center}
\tabcolsep=0.11cm
\par
\begin{tabular}{lcccccccc}
\hline
&  &  &  &  &  &  & \multicolumn{2}{c}{Combined} \\ \hline
& \multicolumn{1}{c}{$\bar{X}$} & \multicolumn{1}{c}{$T^{2}$/$t$} &
\multicolumn{1}{c}{MinP} & \multicolumn{1}{c}{Cl-$T^{2}$} &
\multicolumn{1}{c}{Cl-MinP} & \multicolumn{1}{c}{Cl-Combined} &
\multicolumn{1}{c}{FS} & \multicolumn{1}{c}{SD} \\ \hline
& \multicolumn{8}{c}{$q_{1}=G1$, $q_{2}=G2$} \\ \hline
$T^{2}$ &  & 0.3363 &  &  &  &  & 0.9570 &  \\
Body fat \% & 1.11561 & 0.0372 & 0.1732 & 0.4016 & 0.1732 & 0.1807 & 0.2089
& 0.1732 \\
Pulse rate & 5.64744 & 0.0284 & 0.1578 & 0.4126 & 0.1578 & 0.1642 & 0.1642 &
0.1642 \\
Weight (kg) & 0.01571 & 0.9828 & 0.9962 & 0.9828 & 0.9828 & 0.9828 & 1 &
0.9962 \\
BMI & 0.01585 & 0.9470 & 0.9470 & 0.9470 & 0.9707 & 0.9470 & 1 & 0.9470 \\
Waist (in.) & 0.17537 & 0.7017 & 0.9244 & 0.9444 & 0.9244 & 0.9183 & 0.9995
& 0.9244 \\
Systolic BP & 2.90934 & 0.2703 & 0.7284 & 0.7781 & 0.7284 & 0.7382 & 0.8545
& 0.7284 \\
Diastolic BP & 1.80037 & 0.3071 & 0.6797 & 0.7959 & 0.7284 & 0.7382 & 0.896
& 0.6797 \\ \hline
& \multicolumn{8}{c}{$q_{1}=G1$, $q_{2}=G3$} \\ \hline
$T^{2}$ &  & 0.0033 &  &  &  &  & 0.0798 &  \\
Body fat \% & 2.1886 & 0.0001 & 0.0006 & 0.0033 & 0.0006 & 0.0007 & 0.0007 &
0.0007 \\
Pulse rate & 5.1474 & 0.0619 & 0.2692 & 0.1340 & 0.2692 & 0.2216 & 0.3168 &
0.2692 \\
Weight (kg) & 0.9118 & 0.1316 & 0.3415 & 0.3169 & 0.3415 & 0.3455 & 0.5661 &
0.3415 \\
BMI & 0.3550 & 0.0908 & 0.2500 & 0.2400 & 0.2692 & 0.2503 & 0.4324 & 0.2500
\\
Waist (in.) & 0.7968 & 0.0655 & 0.2382 & 0.1785 & 0.2692 & 0.2380 & 0.3321 &
0.2382 \\
Systolic BP & 3.4474 & 0.1812 & 0.3196 & 0.4032 & 0.3415 & 0.3455 & 0.6931 &
0.3196 \\
Diastolic BP & 0.2885 & 0.8714 & 0.8714 & 0.8714 & 0.8714 & 0.8714 & 1 &
0.8714 \\ \hline
& \multicolumn{8}{c}{$q_{1}=G2$, $q_{2}=G3$} \\ \hline
$T^{2}$ &  & 0.1197 &  &  &  &  & 0.5496 &  \\
Body fat \% & 1.0730 & 0.0027 & 0.0164 & 0.1197 & 0.0164 & 0.0176 & 0.0176 &
0.0176 \\
Pulse rate & 0.5000 & 0.8226 & 0.9711 & 0.9490 & 0.9711 & 0.9711 & 1 & 0.9711
\\
Weight (kg) & 0.9275 & 0.0066 & 0.0337 & 0.1350 & 0.0337 & 0.0356 & 0.0413 &
0.0337 \\
BMI & 0.3391 & 0.0069 & 0.0343 & 0.1350 & 0.0343 & 0.0357 & 0.0428 & 0.0343
\\
Waist (in.) & 0.6214 & 0.0825 & 0.2898 & 0.4447 & 0.2898 & 0.2914 & 0.3999 &
0.2898 \\
Systolic BP & 0.5381 & 0.8465 & 0.8465 & 0.9490 & 0.9711 & 0.9711 & 1 &
0.8465 \\
Diastolic BP & 1.5119 & 0.4106 & 0.7871 & 0.8478 & 0.7871 & 0.7871 & 0.9494
& 0.7871 \\ \hline
\end{tabular}%
\end{center}
\end{table*}

\section{Conclusion}

\label{s:concl}

This paper proposes a combined test for global and multiple hypothesis
testing. The combined test exploits the global power advantages of the
constituent tests while retaining the benefits of the stepdown procedure
used in MinP tests. It also provides a tool for the robust and transparent
evaluation of a global hypothesis when two tests with distinct power
advantages are available for hypothesis testing. A simulation study is
presented to illustrate the power performance of the combined test for
global testing and multiple testing. An empirical application examining the
effects of exercise is presented to illustrate the practical relevance of
the proposed test.

\section*{Acknowledgement}

The author is very grateful to the reviewers for their valuable comments,
which helped improve the presentation of the paper. However, all remaining
errors are the author's own.

\section*{APPENDIX: PROOFS}

\renewcommand{\theequation}{A.\arabic{equation}} \setcounter{equation}{0}

\begin{proof}[Proof of Lemma \protect\ref{th21}]
(i) If Assumption \ref{a0} holds for $l=c$, it follows that for $\varepsilon
>0$
\begin{eqnarray*}
\limsup_{n\rightarrow \infty }E_{P\in \mathbf{P}_{K}}\phi _{c,1}^{(n)}
&=&\limsup_{n\rightarrow \infty }E_{P\in \mathbf{P}_{K}}\mathbf{1(}\hat{p}%
_{c}^{adj}\leq \alpha ) \\
&=&\limsup_{n\rightarrow \infty }G_{c}^{(n)}(\alpha ,P\mathbf{)} \\
&\leq &G_{c}(\alpha +\varepsilon ,P) \\
&\leq &\alpha +\varepsilon ,
\end{eqnarray*}%
where the first inequality follows from Assumption \ref{a0}(i) and (ii), and
the second inequality follows from Assumption \ref{a0}(iii). Since $%
\varepsilon $ can be arbitrarily small, the result follows.

(ii) If Assumption \ref{a0} holds for $l=m,g$, it follows that for $%
\varepsilon >0 $ %
\begin{eqnarray}
&&\limsup_{n\rightarrow \infty }E_{P\in \mathbf{P}_{K}}\phi _{c,2}^{(n)}
\notag \\
&=&\limsup_{n\rightarrow \infty }E_{P\in \mathbf{P}_{K}}\mathbf{1(}\hat{p}%
_{c}^{adj}\leq \alpha )  \notag \\
&\leq &\limsup_{n\rightarrow \infty }E_{P\in \mathbf{P}_{K}}\mathbf{1(}\hat{p%
}_{g}^{adj}\leq \alpha )+\limsup_{n\rightarrow \infty }E_{P\in \mathbf{P}%
_{K}}\mathbf{1(}\hat{p}_{m}^{adj}\leq \alpha )  \notag \\
&&-\liminf_{n\rightarrow \infty }E_{P\in \mathbf{P}_{K}}\mathbf{1(}\hat{p}%
_{g}^{adj}\leq \alpha ,\hat{p}_{m}^{adj}\leq \alpha )+\varepsilon
\label{lmsz1} \\
&\leq &\limsup_{n\rightarrow \infty }E_{P\in \mathbf{P}_{K}}\mathbf{1(}\hat{p%
}_{g}^{adj}\leq \alpha )+\limsup_{n\rightarrow \infty }E_{P\in \mathbf{P}%
_{K}}\mathbf{1(}\hat{p}_{m}^{adj}\leq \alpha )  \notag \\
&=&\limsup_{n\rightarrow \infty }E_{P\in \mathbf{P}_{K}}\mathbf{1(}\hat{p}%
_{g}\leq \alpha /k_{g})+\limsup_{n\rightarrow \infty }E_{P\in \mathbf{P}_{K}}%
\mathbf{1(}\hat{p}_{m}\leq \alpha /k_{m}),  \notag
\end{eqnarray}%
where $k_{g}=2$, $k_{m}=2$, or $k_{g}=(k+1),k_{m}(k+1)/k$ and the last
equality follows from the fact that $\mathbf{1(}\hat{p}_{g}^{adj}\leq \alpha
)=0$ for $\alpha \in (0,1)$ whenever $\hat{p}_{g}^{adj}=1$. The result
follows similarly from the proof in (i) as
\begin{equation*}
\limsup_{n\rightarrow \infty }E_{P\in \mathbf{P}_{K}}\mathbf{1}(\hat{p}%
_{l}\leq \alpha /k_{l})\leq G_{l}(\alpha /k_{l},P)\leq \alpha /k_{l}.
\end{equation*}%
If $\liminf_{n\rightarrow \infty }E_{P\in \mathbf{P}_{K}}\phi
_{l}^{(n)}(1-\phi _{l^{\prime }}^{(n)})>0$ for $l,l^{\prime }\in \{g,m\}$
and $l\neq l^{\prime }$, then (\ref{lmsz1}) may be rewritten as%
\begin{eqnarray*}
&&\limsup_{n\rightarrow \infty }E_{P\in \mathbf{P}_{K}}\phi _{c,2}^{(n)} \\
&\leq &\limsup_{n\rightarrow \infty }E_{P\in \mathbf{P}_{K}}\phi
_{g}^{(n)}+\limsup_{n\rightarrow \infty }E_{P\in \mathbf{P}_{K}}\phi
_{m}^{(n)}-\liminf_{n\rightarrow \infty }E_{P\in \mathbf{P}_{K}}\phi
_{g}^{(n)}\phi _{m}^{(n)}+\varepsilon \\
&<&\limsup_{n\rightarrow \infty }E_{P\in \mathbf{P}_{K}}\phi
_{g}^{(n)}+\limsup_{n\rightarrow \infty }E_{P\in \mathbf{P}_{K}}\phi
_{m}^{(n)} \\
&\leq &\alpha .
\end{eqnarray*}
\end{proof}

\begin{proof}[Proof of Theorem \protect\ref{adth}]
We prove the asymptotic admissibility using a contradiction argument. Suppose the combined
test $\phi _{c}^{(n)}$ is not asymptotically $d$-admissible. It follows from
(\ref{eqad1}) and (\ref{eqad2}) that there exists some other test $\phi
^{(n)} $ such that%
\begin{equation}
\limsup_{n\rightarrow \infty }E_{P}(\phi _{c}^{(n)})\leq
\liminf_{n\rightarrow \infty }E_{P}(\phi ^{(n)}),\qquad \forall P\in
\{P_{n}:P_{n}\in \mathbf{P}_{n,K}^{\prime }\}  \label{thad10}
\end{equation}%
\begin{equation}
\liminf_{n\rightarrow \infty }E_{P}(\phi _{c}^{(n)})\geq
\limsup_{n\rightarrow \infty }E_{P}(\phi ^{(n)}),\qquad \forall P\in \mathbf{%
P}_{K}  \label{thad20}
\end{equation}%
hold with at least one strict inequality.

By the construction of the combined test, the event that $\phi _{c}^{(n)}=1$
occurs implies that at least one of the two events--$\tilde{\phi}_{g}^{(n)}=%
\mathbf{1}(\hat{p}_{g}^{adj}\leq \alpha )=1$ and $\tilde{\phi}_{m}^{(n)}=%
\mathbf{1}(\hat{p}_{m}^{adj}\leq \alpha )=1$ must occur. That is,
\begin{equation}
\{\phi _{c}^{(n)}=1\}=\{\tilde{\phi}_{g}^{(n)}=1\}\cup \{\tilde{\phi}%
_{m}^{(n)}=1\},  \label{thad3}
\end{equation}%
\begin{equation}
\{\phi _{c}^{(n)}=0\}=\{\tilde{\phi}_{g}^{(n)}=0\}\cap \{\tilde{\phi}%
_{m}^{(n)}=0\}.  \label{thad31}
\end{equation}%

From (\ref{thad10}) and (\ref{thad3}) we have for $l=g,m$,
\begin{equation}
\limsup_{n\rightarrow \infty }E_{P}(\tilde{\phi}_{l}^{(n)})\leq
\liminf_{n\rightarrow \infty }E_{P}(\phi ^{(n)}),\qquad \forall P\in
\{P_{n}:P_{n}\in \mathbf{P}_{n,K}^{\prime }\}.  \label{thad11}
\end{equation}%
From (\ref{thad20}) and (\ref{thad31}) we have for $l=g,m$,
\begin{equation}
\liminf_{n\rightarrow \infty }E_{P}(\tilde{\phi}_{l}^{(n)})\geq
\limsup_{n\rightarrow \infty }E_{P}(\phi ^{(n)}),\qquad \forall P\in \mathbf{%
P}_{K}.  \label{thad21}
\end{equation}%
Therefore, (\ref{thad11}) and (\ref{thad21}) must hold with at least one
strict inequality. Consequently, this suggests that $\tilde{\phi}_{l}^{(n)}$%
, $l=g,m$, is not $d$-admissible. Next, we shall show that this implies $%
\phi _{l}^{(n)}$, $l=g,m$, is not $d$-admissible, which contradicts the
assumption and hence proves the asymptotic $d$-admissibility of $\phi
_{c}^{(n)}$.

If $\hat{p}_{l}^{adj}\leq \alpha $, $l=g,m$, then $\hat{p}_{l}\leq
(G_{c}^{(n)})^{-1}(\alpha ,P\in \mathbf{P}_{K})$, where $(G_{c}^{(n)})^{-1}$
is the inverse function of $G_{c}^{(n)}$, or $\hat{p}_{l}\leq \alpha /b$,
where $b=2$, or%
\begin{equation*}
b=\left\{
\begin{array}{c}
k+1,\quad \text{when }l=g, \\
\frac{k+1}{k},\quad \text{\ \ when }l=m.%
\end{array}%
\right.
\end{equation*}
Let $\varphi ^{(n)}=\mathbf{1}(\hat{p}_{l}\leq \tau (\alpha ))$, where $\tau
(\alpha )=(G_{c}^{(n)})^{-1}(\alpha ,P\in \mathbf{P}_{K})$, or $\alpha /b$,
be the test corresponding to $\phi ^{(n)}$ in (\ref{thad11}) and (\ref%
{thad21}). By Assumption \ref{a0}(i)-(ii) for $l=c$ we have%
\begin{equation*}
(G_{c}^{(n)})^{-1}(\alpha ,P\in \mathbf{P}_{K})\rightarrow G_{c}^{-1}(\alpha
,P\in \mathbf{P}_{K}).
\end{equation*}%
Therefore, it follows from (\ref{thad11}) and (\ref{thad21}) that there
exists a test $\varphi ^{(n)}$ such that%
\begin{equation*}
\limsup_{n\rightarrow \infty }E_{P}(\phi _{l}^{(n)})\leq
\liminf_{n\rightarrow \infty }E_{P}(\varphi ^{(n)}),\qquad \forall P\in
\{P_{n}:P_{n}\in \mathbf{P}_{n,K}^{\prime }\},
\end{equation*}%
\begin{equation*}
\liminf_{n\rightarrow \infty }E_{P}(\phi _{l}^{(n)})\geq
\limsup_{n\rightarrow \infty }E_{P}(\varphi ^{(n)}),\qquad \forall P\in
\mathbf{P}_{K},
\end{equation*}%
hold with at least one strict inequality. This suggests $\phi _{l}^{(n)}$, $l=g,m$, is not $d$-admissible.
\end{proof}

\begin{proof}[Proof of Theorem \protect\ref{adthb}]
By Theorem \ref{adth} $\phi _{c}^{(n)}$ is $d$-admissible meaning there
cannot exist any other test $\phi ^{(n)}$ such that (\ref{thad10}) and (\ref%
{thad20}) hold with at least one strict inequality. If $\lim_{n\rightarrow
\infty }E_{P}(\phi _{c}^{(n)})=\alpha $, then (\ref{thad20}) must hold if $%
\phi ^{(n)}$ is an asymptotically level-$\alpha $ test. Therefore, (\ref{thad10}%
) cannot hold. That is, there cannot exist any other level-$\alpha $ test $%
\phi ^{(n)}$ such that (\ref{thad10}) holds.
\end{proof}

\begin{proof}[Proof of Theorem \protect\ref{th27}]
It follows that
\begin{eqnarray}
&&\max (\tilde{\phi}_{l}^{(n)},l=g,m)  \notag \\
&=&\tilde{\phi}_{l}^{(n)}\mathbf{1}(\tilde{\phi}_{l}^{(n)}\geq \tilde{\phi}%
_{l^{\prime }}^{(n)})+\tilde{\phi}_{l^{\prime }}^{(n)}\mathbf{1}(\tilde{\phi}%
_{l^{\prime }}^{(n)}>\tilde{\phi}_{l}^{(n)})  \notag \\
&=&\tilde{\phi}_{l}^{(n)}\mathbf{1}(\tilde{\phi}_{l}^{(n)}\geq \tilde{\phi}%
_{l^{\prime }}^{(n)})+\tilde{\phi}_{l}^{(n)}\mathbf{1}(\tilde{\phi}%
_{l^{\prime }}^{(n)}>\tilde{\phi}_{l}^{(n)})+(\tilde{\phi}_{l^{\prime
}}^{(n)}-\tilde{\phi}_{l}^{(n)})\mathbf{1}(\tilde{\phi}_{l^{\prime }}^{(n)}>%
\tilde{\phi}_{l}^{(n)})  \notag \\
&=&\tilde{\phi}_{l}^{(n)}+(\tilde{\phi}_{l^{\prime }}^{(n)}-\tilde{\phi}%
_{l}^{(n)})\mathbf{1}(\tilde{\phi}_{l^{\prime }}^{(n)}>\tilde{\phi}%
_{l}^{(n)}).  \label{thad4}
\end{eqnarray}%
By (\ref{thad3}) we can rewrite
\begin{equation}
\phi _{c}^{(n)}=\max (\tilde{\phi}_{l}^{(n)},l=g,m).  \label{thad40}
\end{equation}%
It then follows from (\ref{thad4}) and (\ref{thad40}) that%
\begin{equation}
\phi _{c}^{(n)}=\tilde{\phi}_{l}^{(n)}+(\tilde{\phi}_{l^{\prime }}^{(n)}-%
\tilde{\phi}_{l}^{(n)})\mathbf{1}(\tilde{\phi}_{l^{\prime }}^{(n)}>\tilde{%
\phi}_{l}^{(n)}),  \label{thad70}
\end{equation}%
which gives rise to%
\begin{equation*}
\phi _{c}^{(n)}-\phi _{l}^{(n)}=(\tilde{\phi}_{l^{\prime }}^{(n)}-\tilde{\phi%
}_{l}^{(n)})\mathbf{1}(\tilde{\phi}_{l^{\prime }}^{(n)}>\tilde{\phi}%
_{l}^{(n)})-(\phi _{l}^{(n)}-\tilde{\phi}_{l}^{(n)}).
\end{equation*}%
It follows that for $\varepsilon >0$%
\begin{eqnarray}
&&\liminf_{n\rightarrow \infty }E_{P\in \mathbf{P}_{n,K}^{\prime }}\phi
_{c}^{(n)}-\limsup_{n\rightarrow \infty }E_{P\in \mathbf{P}_{n,K}^{\prime
}}\phi _{l}^{(n)}  \notag \\
&\geq &\liminf_{n\rightarrow \infty }E_{P\in \mathbf{P}_{n,K}^{\prime
}}(\phi _{c}^{(n)}-\phi _{l}^{(n)})-\varepsilon  \label{thad7} \\
&=&\liminf_{n\rightarrow \infty }E_{P\in \mathbf{P}_{n,K}^{\prime }}\{(%
\tilde{\phi}_{l^{\prime }}^{(n)}-\tilde{\phi}_{l}^{(n)})\mathbf{1}(\tilde{%
\phi}_{l^{\prime }}^{(n)}>\tilde{\phi}_{l}^{(n)})-(\phi _{l}^{(n)}-\tilde{%
\phi}_{l}^{(n)})\}-\varepsilon .  \notag
\end{eqnarray}%
Since $\varepsilon $ can be arbitrarily small, the sufficiency result follows.

We now turn to proving the necessity result by contradiction. Suppose that (\ref%
{eqga2}) does not hold, then this implies that
\begin{equation*}
\liminf_{n\rightarrow \infty }E_{P\in \mathbf{P}_{n,K}^{\prime }}\{(\tilde{%
\phi}_{l^{\prime }}^{(n)}-\tilde{\phi}_{l}^{(n)})\mathbf{1}(\tilde{\phi}%
_{l^{\prime }}^{(n)}>\tilde{\phi}_{l}^{(n)})-(\phi _{l}^{(n)}-\tilde{\phi}%
_{l}^{(n)})\}\leq 0.
\end{equation*}%
Similar to (\ref{thad7}), it follows that for $\varepsilon _{1},\varepsilon
_{2}>0$
\begin{eqnarray*}
&&\limsup_{n\rightarrow \infty }E_{P\in \mathbf{P}_{n,K}^{\prime }}\phi
_{c}^{(n)}-\liminf_{n\rightarrow \infty }E_{P\in \mathbf{P}_{n,K}^{\prime
}}\phi _{l}^{(n)} \\
&\leq &\limsup_{n\rightarrow \infty }E_{P\in \mathbf{P}_{n,K}^{\prime
}}(\phi _{c}^{(n)}-\phi _{l}^{(n)})+\varepsilon _{1} \\
&\leq &\liminf_{n\rightarrow \infty }E_{P\in \mathbf{P}_{n,K}^{\prime
}}(\phi _{c}^{(n)}-\phi _{l}^{(n)})+\varepsilon _{1}+\varepsilon _{2} \\
&=&\liminf_{n\rightarrow \infty }E_{P\in \mathbf{P}_{n,K}^{\prime }}\{(%
\tilde{\phi}_{l^{\prime }}^{(n)}-\tilde{\phi}_{l}^{(n)})\mathbf{1}(\tilde{%
\phi}_{l^{\prime }}^{(n)}>\tilde{\phi}_{l}^{(n)})-(\phi _{l}^{(n)}-\tilde{%
\phi}_{l}^{(n)})\}+\varepsilon _{1}+\varepsilon _{2}.
\end{eqnarray*}%
Since $\varepsilon _{1}$ and $\varepsilon _{2}$ can be arbitrarily small, it
follows that
\begin{equation*}
\limsup_{n\rightarrow \infty }E_{P\in \mathbf{P}_{n,K}^{\prime }}\phi
_{c}^{(n)}\leq \liminf_{n\rightarrow \infty }E_{P\in \mathbf{P}%
_{n,K}^{\prime }}\phi _{l}^{(n)},
\end{equation*}%
which implies that $\phi _{c}^{(n)}$ does not improve the global power of $%
\phi _{l}^{(n)}$. This completes the proof of necessity.
\end{proof}

\begin{proof}[Proof of Theorem \protect\ref{thblndp}]
The result in (\ref{eqthblndp1a}) follows easily from Theorem \ref{th27}. We
now prove the result in (\ref{eqthblndp1b}). It follows from (\ref{thad70})
that
\begin{eqnarray*}
\phi _{c}^{(n)}-\phi _{l^{\prime }}^{(n)} &=&(\tilde{\phi}_{l^{\prime
}}^{(n)}-\tilde{\phi}_{l}^{(n)})\mathbf{1}(\tilde{\phi}_{l^{\prime }}^{(n)}>%
\tilde{\phi}_{l}^{(n)})+\tilde{\phi}_{l}^{(n)}-\phi _{l^{\prime }}^{(n)} \\
&=&\psi ^{(n)}-(\phi _{l^{\prime }}^{(n)}-\phi _{l}^{(n)}).
\end{eqnarray*}%
Therefore, if
\begin{equation*}
\limsup_{n\rightarrow \infty }E_{P\in \mathbf{P}_{n,K}^{\prime }}\psi
^{(n)}\leq \liminf_{n\rightarrow \infty }E_{P\in \mathbf{P}_{n,K}^{\prime
}}(\phi _{l^{\prime }}^{(n)}-\phi _{l}^{(n)}),
\end{equation*}%
(\ref{eqthblndp1b}) follows.
\end{proof}

\begin{proof}[Proof of Theorem \protect\ref{th23}]
Let the subscripts $m$ and $c$ in FWER indicate controls under the
MinP test $\phi _{m}^{(n)}$ and the combined test $\phi _{c}^{(n)}$,
respectively. If the MinP test used in the combined procedure controls
the FWER, it implies that for $K_{\ast }\subseteq K_{i}\subseteq K$,
\begin{eqnarray*}
\limsup_{n\rightarrow \infty } \text{FWER}_{m} &=&\limsup_{n\rightarrow \infty
}\Pr_{P\in \mathbf{P}_{K_{\ast }}}(\text{reject at least one }H_{i} for i\in
K_{\ast }\text{, by }\phi _{m}^{(n)}) \\
&=&\limsup_{n\rightarrow \infty }\Pr_{P\in \mathbf{P}_{K_{\ast }}}\{\min (%
\hat{p}_{m,i}^{adj},i\in K_{\ast })\leq \alpha \} \\
&\leq &\limsup_{n\rightarrow \infty }\Pr_{P\in \mathbf{P}_{K_{i}}}\{\min (%
\hat{p}_{m,i}^{adj},i\in K_{i})\leq \alpha \} \\
&=&\limsup_{n\rightarrow \infty }\Pr_{P\in \mathbf{P}_{K_{i}}}\{\hat{p}%
_{m,(i)}^{adj}\leq \alpha \} \\
&\leq &\alpha .
\end{eqnarray*}%
That is, the MinP test controls the FWER at any $K_{i}$ that contains the
true null set $K_{\ast }$. This holds at the first step of the stepdown
procedure when $K_{i}=K$;
\begin{equation}
\limsup_{n\rightarrow \infty }\Pr_{P\in \mathbf{P}_{K_{\ast }}}\{\min (\hat{p%
}_{i}^{adj},i\in K_{\ast })\leq \alpha )\}\leq \limsup_{n\rightarrow \infty
}\Pr_{P\in \mathbf{P}_{K}}(\hat{p}_{m,(1)}^{adj}\leq \alpha )\leq \alpha .
\label{eqth23b}
\end{equation}

The combined test and the MinP test differ only in the
first step of the stepdown procedure for multiple testing.
Therefore, to prove that the combined test controls the FWER, it suffices to
show that the combined test controls the FWER at the first step.%
\begin{eqnarray}
\limsup_{n\rightarrow \infty }\text{FWER}_{c} &=&\limsup_{n\rightarrow \infty
}\Pr_{P\in \mathbf{P}_{K_{\ast }}}(\text{reject at least one }H_{i},i\in
K_{\ast }\text{, by }\phi _{c}^{(n)})  \notag \\
&=&\limsup_{n\rightarrow \infty }\Pr_{P\in \mathbf{P}_{K_{\ast }}}\{\min (%
\hat{p}_{i}^{adj},i\in K_{\ast })\leq \alpha \}  \notag \\
&\leq &\limsup_{n\rightarrow \infty }\Pr_{P\in \mathbf{P}_{K_{\ast }}}\{\min
(\hat{p}_{i}^{adj},i\in K_{\ast })\leq \alpha \},  \label{eqth23c}
\end{eqnarray}%
where $\hat{p}_{i}^{adj}=G_{c}^{(n)}(\hat{p}_{i},P\in \mathbf{P}_{K})$ and
the inequality follows because $\hat{p}_{g}\wedge \hat{p}_{m}\leq \hat{p}%
_{m} $ implying
\begin{equation*}
\hat{p}_{i}^{adj}=G_{c}^{(n)}(\hat{p}_{i},P\in \mathbf{P}_{K})\geq
G_{m,K}^{(n)}(\hat{p}_{i},P\in \mathbf{P}_{K}\mathbf{)=}\hat{p}_{m,i}^{adj},
\end{equation*}%
hence,
\begin{equation*}
\{\min (\hat{p}_{i}^{adj},i\in K_{\ast })\leq \alpha \}\subseteq \{\min (%
\hat{p}_{m,i}^{adj},i\in K_{\ast })\leq \alpha \}.
\end{equation*}%
It follows from (\ref{eqth23b}) and (\ref{eqth23c}) that $%
\limsup_{n\rightarrow \infty }\text{FWER}_{c}\leq \alpha $.
\end{proof}

\bigskip
\bibliographystyle{model2-names}

\section*{Additional results from the Monte Carlo study}
\setcounter{table}{0}
\renewcommand{\thetable}{S\arabic{table}}

This supplemental file presents additional Monte Carlo study results for a moderately hogh dimension $k$. Since the computational cost of closed testing increases exponentially with $k$, we have excluded those computations from this study. The Monte Carlo study results presented in this supplemental file further support the conclusions reported for the lower-dimensional scenarios discussed in the main text.

\begin{table*}[tbp]
\caption{The estimated sizes and powers in percentages in global testing, and
the estimated FWER (in percentages) and the ANCR (average number of correctly rejected false $H_{i}$) in multiple
testing procedures based on bootstrap with $n=100$.}
\label{tbs1}
\begin{center}
\tabcolsep=0.11cm
\par
\begin{tabular}{lccccccccc}
\hline
\multicolumn{1}{c}{$\mu$} & \multicolumn{3}{c}{Combined} &
\multicolumn{3}{c}{MinP} & \multicolumn{3}{c}{$T^{2}$} \\ \hline
& \multicolumn{1}{c}{$H_{K}$} & \multicolumn{1}{c}{FWER} &
\multicolumn{1}{c}{ANCR} & \multicolumn{1}{c}{$H_{K}$} & \multicolumn{1}{c}{
FWER} & \multicolumn{1}{c}{ANCR} & \multicolumn{1}{c}{$H_{K}$} &
\multicolumn{1}{c}{FWER} & \multicolumn{1}{c}{ANCR} \\ \hline
& \multicolumn{9}{c}{$\rho_{ij}=-0.25, i\neq j$} \\ \hline
$\mathbbm{1}_{4,0}$ & 4.85 & 4.50 & 0 & 4.85 & 4.85 & 0 & 4.65 & 2.75 & 0 \\
0.22$\mathbbm{1}_{4,1}$ & 47.60 & 2.55 & 0.37 & 40.35 & 2.70 & 0.38 & 55.80
& 3.45 & 0.31 \\
0.13$\mathbbm{1}_{4,2}$ & 41.95 & 1.95 & 0.22 & 24.65 & 2.10 & 0.23 & 59.10
& 2.55 & 0.20 \\
0.1$\mathbbm{1}_{4,3}$ & 49.45 & 1.05 & 0.19 & 20.75 & 1.10 & 0.20 & 69.50 &
1.10 & 0.19 \\
0.07$\mathbbm{1}_{4,4}$ & 36.30 & 0 & 0.13 & 13.40 & 0 & 0.13 & 56.05 & 0 &
0.12 \\ \hline
& \multicolumn{9}{c}{$\rho_{ij}=0, i\neq j$} \\ \hline
$\mathbbm{1}_{4,0}$ & 5.25 & 5.15 & 0 & 5.30 & 5.30 & 0 & 5.40 & 3.30 & 0 \\
0.25$\mathbbm{1}_{4,1}$ & 51.25 & 2.55 & 0.48 & 51.00 & 2.60 & 0.48 & 45.55
& 4.00 & 0.38 \\
0.2$\mathbbm{1}_{4,2}$ & 54.85 & 2.00 & 0.51 & 53.80 & 2.05 & 0.52 & 56.30 &
3.85 & 0.55 \\
0.15$\mathbbm{1}_{4,3}$ & 42.10 & 1.15 & 0.39 & 41.00 & 1.15 & 0.40 & 48.20
& 1.60 & 0.45 \\
0.14$\mathbbm{1}_{4,4}$ & 47.25 & 0 & 0.43 & 43.85 & 0 & 0.44 & 56.55 & 0 &
0.61 \\ \hline
& \multicolumn{9}{c}{$\rho_{ij}=0.5, i\neq j$} \\ \hline
$\mathbbm{1}_{4,0}$ & 5.10 & 4.55 & 0 & 4.70 & 4.70 & 0 & 5.80 & 2.75 & 0 \\
0.22$\mathbbm{1}_{4,1}$ & 49.15 & 2.20 & 0.38 & 41.80 & 2.50 & 0.39 & 55.35
& 3.55 & 0.37 \\
0.17$\mathbbm{1}_{4,2}$ & 43.50 & 1.85 & 0.32 & 35.60 & 1.85 & 0.34 & 50.15
& 3.50 & 0.38 \\
0.17$\mathbbm{1}_{4,3}$ & 48.65 & 0.75 & 0.42 & 44.70 & 0.90 & 0.44 & 51.85
& 2.75 & 0.61 \\
0.17$\mathbbm{1}_{4,4}$ & 47.55 & 0 & 0.47 & 48.90 & 0 & 0.49 & 35.50 & 0 &
0.78 \\ \hline
& \multicolumn{9}{c}{$\rho_{ij}=0.9, i\neq j$} \\ \hline
$\mathbbm{1}_{4,0}$ & 4.75 & 3.55 & 0 & 4.80 & 4.80 & 0 & 4.35 & 1.15 & 0 \\
0.11$\mathbbm{1}_{4,1}$ & 50 & 2.05 & 0.11 & 15.60 & 2.60 & 0.13 & 62.45 &
1.55 & 0.16 \\
0.09$\mathbbm{1}_{4,2}$ & 46.30 & 1.25 & 0.11 & 14.05 & 1.75 & 0.12 & 59.85
& 2.65 & 0.14 \\
0.1$\mathbbm{1}_{4,3}$ & 46.15 & 1.25 & 0.13 & 16.95 & 1.55 & 0.15 & 58.90 &
4.20 & 0.19 \\
0.2$\mathbbm{1}_{4,4}$ & 50.15 & 0 & 0.50 & 53.95 & 0 & 0.54 & 34.25 & 0 &
1.01 \\ \hline
& \multicolumn{9}{c}{$\rho_{ij}=0.5^{\left\vert i-j\right\vert }$} \\ \hline
$\mathbbm{1}_{4,0}$ & 6.05 & 5.80 & 0 & 6.10 & 6.10 & 0 & 4.70 & 2.65 & 0 \\
0.24$\mathbbm{1}_{4,1}$ & 53.20 & 2.85 & 0.44 & 48.45 & 3.10 & 0.45 & 56.30
& 3.05 & 0.39 \\
0.2$\mathbbm{1}_{4,2}$ & 49.85 & 1.70 & 0.45 & 48.40 & 1.80 & 0.47 & 50.80 &
4.00 & 0.57 \\
0.18$\mathbbm{1}_{4,3}$ & 51.60 & 0.90 & 0.49 & 50.80 & 0.95 & 0.50 & 49.30
& 3.10 & 0.70 \\
0.17$\mathbbm{1}_{4,4}$ & 49.55 & 0 & 0.49 & 50.40 & 0 & 0.50 & 40.25 & 0 &
0.78 \\ \hline
& \multicolumn{9}{c}{$\rho_{ij}=(-0.5)^{\left\vert i-j\right\vert }$} \\
\hline
$\mathbbm{1}_{4,0}$ & 5.70 & 5.20 & 0 & 5.65 & 5.65 & 0 & 5.90 & 3.00 & 0 \\
0.24$\mathbbm{1}_{4,1}$ & 52.55 & 1.95 & 0.46 & 49.60 & 2.25 & 0.47 & 56.45
& 2.70 & 0.40 \\
0.13$\mathbbm{1}_{4,2}$ & 37.00 & 2.00 & 0.22 & 24.80 & 2.10 & 0.23 & 52.15
& 2.55 & 0.21 \\
0.11$\mathbbm{1}_{4,3}$ & 46.90 & 0.60 & 0.24 & 25.65 & 0.70 & 0.25 & 64.30
& 1.65 & 0.27 \\
0.09$\mathbbm{1}_{4,4}$ & 40.65 & 0 & 0.20 & 21.35 & 0 & 0.21 & 58.60 & 0 &
0.23 \\ \hline
\end{tabular}%
\end{center}
\end{table*}

\begin{table*}[tbp]
\caption{The estimated sizes and powers (in percentages) in global testing, and
the estimated FWER (in percentages) and the ANCR (average number of correctly rejected false $H_{i}$) in multiple
testing procedures based on the limiting normal distribution with $n=100$.}
\label{tb3}
\begin{center}
\tabcolsep=0.11cm
\par
\begin{tabular}{lccccccccc}
\hline
\multicolumn{1}{c}{$\mu$} & \multicolumn{3}{c}{Combined} &
\multicolumn{3}{c}{MinP} & \multicolumn{3}{c}{$T^{2}$} \\ \hline
& \multicolumn{1}{c}{$H_{K}$} & \multicolumn{1}{c}{FWER} &
\multicolumn{1}{c}{ANCR} & \multicolumn{1}{c}{$H_{K}$} & \multicolumn{1}{c}{
FWER} & \multicolumn{1}{c}{ANCR} & \multicolumn{1}{c}{$H_{K}$} &
\multicolumn{1}{c}{FWER} & \multicolumn{1}{c}{ANCR} \\ \hline
& \multicolumn{9}{c}{$\rho_{ij}=-0.25, i\neq j$} \\ \hline
$\mathbbm{1}_{4,0}$ & 5.85 & 5.30 & 0 & 5.35 & 5.35 & 0 & 5.05 & 4.70 & 0 \\
0.22$\mathbbm{1}_{4,1}$ & 52.00 & 5.20 & 0.39 & 43.15 & 5.45 & 0.40 & 58.35
& 8.30 & 0.43 \\
0.13$\mathbbm{1}_{4,2}$ & 48.35 & 3.35 & 0.25 & 27.50 & 3.55 & 0.26 & 59.45
& 5.80 & 0.34 \\
0.1$\mathbbm{1}_{4,3}$ & 56.80 & 1.55 & 0.24 & 24.35 & 1.80 & 0.24 & 68.70 &
2.90 & 0.33 \\
0.07$\mathbbm{1}_{4,4}$ & 45.05 & 0 & 0.17 & 16.00 & 0 & 0.17 & 57.65 & 0 &
0.25 \\ \hline
& \multicolumn{9}{c}{$\rho_{ij}=0, i\neq j$} \\ \hline
$\mathbbm{1}_{4,0}$ & 5.70 & 5.40 & 0 & 5.45 & 5.45 & 0 & 4.55 & 3.70 & 0 \\
0.25$\mathbbm{1}_{4,1}$ & 52.85 & 5.00 & 0.50 & 53.05 & 5.15 & 0.51 & 47.75
& 4.15 & 0.44 \\
0.2$\mathbbm{1}_{4,2}$ & 57.20 & 3.90 & 0.67 & 54.75 & 3.85 & 0.67 & 57.35 &
4.10 & 0.60 \\
0.15$\mathbbm{1}_{4,3}$ & 49.60 & 1.80 & 0.56 & 45.40 & 1.75 & 0.56 & 52.80
& 2.75 & 0.56 \\
0.14$\mathbbm{1}_{4,4}$ & 50.70 & 0 & 0.60 & 45.75 & 0 & 0.61 & 56.50 & 0 &
0.68 \\ \hline
& \multicolumn{9}{c}{$\rho_{ij}=0.5, i\neq j$} \\ \hline
$\mathbbm{1}_{4,0}$ & 5.45 & 4.90 & 0 & 5.50 & 5.50 & 0 & 4.70 & 2.75 & 0 \\
0.22$\mathbbm{1}_{4,1}$ & 50.95 & 5.00 & 0.36 & 40.75 & 5.45 & 0.38 & 55.25
& 3.40 & 0.31 \\
0.17$\mathbbm{1}_{4,2}$ & 46.80 & 3.40 & 0.47 & 37.90 & 3.55 & 0.49 & 52.20
& 2.45 & 0.38 \\
0.17$\mathbbm{1}_{4,3}$ & 50.65 & 2.45 & 0.75 & 45.20 & 2.45 & 0.77 & 51.00
& 1.85 & 0.63 \\
0.17$\mathbbm{1}_{4,4}$ & 48.35 & 0 & 1.08 & 50.85 & 0 & 1.12 & 34.90 & 0 &
0.88 \\ \hline
& \multicolumn{9}{c}{$\rho_{ij}=0.9, i\neq j$} \\ \hline
$\mathbbm{1}_{4,0}$ & 5.45 & 4.10 & 0 & 5.30 & 5.30 & 0 & 5.00 & 2.10 & 0 \\
0.11$\mathbbm{1}_{4,1}$ & 56.10 & 4.90 & 0.11 & 15.05 & 5.35 & 0.13 & 63.85
& 2.55 & 0.16 \\
0.09$\mathbbm{1}_{4,2}$ & 51.50 & 3.40 & 0.17 & 14.20 & 3.90 & 0.20 & 59.75
& 2.50 & 0.15 \\
0.1$\mathbbm{1}_{4,3}$ & 51.40 & 3.85 & 0.34 & 17.95 & 4.15 & 0.37 & 57.50 &
4.50 & 0.22 \\
0.2$\mathbbm{1}_{4,4}$ & 49.50 & 0 & 1.70 & 54.05 & 0 & 1.78 & 32.20 & 0 &
1.06 \\ \hline
& \multicolumn{9}{c}{$\rho_{ij}=0.5^{\left\vert i-j\right\vert }$} \\ \hline
$\mathbbm{1}_{4,0}$ & 6.60 & 5.85 & 0 & 6.10 & 6.10 & 0 & 5.45 & 3.00 & 0 \\
0.24$\mathbbm{1}_{4,1}$ & 56.65 & 4.35 & 0.49 & 51.80 & 4.60 & 0.50 & 58.25
& 3.05 & 0.41 \\
0.2$\mathbbm{1}_{4,2}$ & 52.80 & 4.40 & 0.68 & 49.50 & 4.45 & 0.68 & 49.25 &
3.65 & 0.52 \\
0.18$\mathbbm{1}_{4,3}$ & 52.20 & 2.30 & 0.83 & 49.55 & 2.35 & 0.84 & 48.40
& 2.05 & 0.65 \\
0.17$\mathbbm{1}_{4,4}$ & 51.95 & 0 & 1.06 & 52.40 & 0 & 1.07 & 43.65 & 0 &
0.92 \\ \hline
& \multicolumn{9}{c}{$\rho_{ij}=(-0.5)^{\left\vert i-j\right\vert }$} \\
\hline
$\mathbbm{1}_{4,0}$ & 5.55 & 5.00 & 0 & 5.30 & 5.30 & 0 & 4.95 & 3.60 & 0 \\
0.24$\mathbbm{1}_{4,1}$ & 56.05 & 5.10 & 0.48 & 50.50 & 5.20 & 0.48 & 56.20
& 7.50 & 0.44 \\
0.13$\mathbbm{1}_{4,2}$ & 42.95 & 3.15 & 0.24 & 27.25 & 3.70 & 0.25 & 54.00
& 5.15 & 0.26 \\
0.11$\mathbbm{1}_{4,3}$ & 49.10 & 1.55 & 0.25 & 24.65 & 1.55 & 0.26 & 61.55
& 1.75 & 0.31 \\
0.09$\mathbbm{1}_{4,4}$ & 47.80 & 0 & 0.25 & 23.90 & 0 & 0.26 & 58.25 & 0 &
0.34 \\ \hline
\end{tabular}%
\end{center}
\end{table*}

\begin{table*}[tbp]
\caption{The estimated sizes and powers (in percentages) in global testing, and
the estimated FWER (in percentages) and the ANCR (average number of correctly rejected false $H_{i}$) in multiple
testing procedures based on the limiting normal distribution with $n=200$.}
\label{tb4}
\begin{center}
\tabcolsep=0.11cm
\par
\begin{tabular}{lccccccccc}
\hline
\multicolumn{1}{c}{$\mu$} & \multicolumn{3}{c}{Combined} &
\multicolumn{3}{c}{MinP} & \multicolumn{3}{c}{$T^{2}$} \\ \hline
& \multicolumn{1}{c}{$H_{K}$} & \multicolumn{1}{c}{FWER} &
\multicolumn{1}{c}{ANCR} & \multicolumn{1}{c}{$H_{K}$} & \multicolumn{1}{c}{
FWER} & \multicolumn{1}{c}{ANCR} & \multicolumn{1}{c}{$H_{K}$} &
\multicolumn{1}{c}{FWER} & \multicolumn{1}{c}{ANCR} \\ \hline
& \multicolumn{9}{c}{$\rho_{ij}=-0.25, i\neq j$} \\ \hline
$\mathbbm{1}_{4,0}$ & 5.80 & 5.45 & 0 & 5.75 & 5.75 & 0 & 5.40 & 5.00 & 0 \\
0.22$\mathbbm{1}_{4,1}$ & 85.05 & 4.60 & 0.73 & 75.30 & 4.90 & 0.74 & 89.90
& 8.40 & 0.78 \\
0.13$\mathbbm{1}_{4,2}$ & 81.35 & 3.40 & 0.52 & 49.10 & 3.60 & 0.53 & 89.70
& 6.80 & 0.67 \\
0.1$\mathbbm{1}_{4,3}$ & 89.15 & 1.75 & 0.46 & 43.25 & 1.75 & 0.47 & 95.00 &
3.70 & 0.67 \\
0.07$\mathbbm{1}_{4,4}$ & 81.65 & 0 & 0.30 & 28.75 & 0 & 0.30 & 90.65 & 0 &
0.45 \\ \hline
& \multicolumn{9}{c}{$\rho_{ij}=0, i\neq j$} \\ \hline
$\mathbbm{1}_{4,0}$ & 4.25 & 4.25 & 0 & 4.35 & 4.35 & 0 & 3.95 & 2.55 & 0 \\
0.25$\mathbbm{1}_{4,1}$ & 85.10 & 4.70 & 0.84 & 84.95 & 4.75 & 0.84 & 80.05
& 3.65 & 0.78 \\
0.2$\mathbbm{1}_{4,2}$ & 88.10 & 5.05 & 1.32 & 86.95 & 5.05 & 1.32 & 91.10 &
5.35 & 1.25 \\
0.15$\mathbbm{1}_{4,3}$ & 78.50 & 2.30 & 1.18 & 74.45 & 2.30 & 1.17 & 83.75
& 3.75 & 1.21 \\
0.14$\mathbbm{1}_{4,4}$ & 84.35 & 0 & 1.43 & 77.75 & 0 & 1.44 & 90.15 & 0 &
1.68 \\ \hline
& \multicolumn{9}{c}{$\rho_{ij}=0.5, i\neq j$} \\ \hline
$\mathbbm{1}_{4,0}$ & 4.80 & 4.00 & 0 & 4.35 & 4.35 & 0 & 4.40 & 2.45 & 0 \\
0.22$\mathbbm{1}_{4,1}$ & 85.45 & 4.60 & 0.74 & 76.80 & 4.85 & 0.75 & 89.40
& 2.65 & 0.68 \\
0.17$\mathbbm{1}_{4,2}$ & 78.60 & 3.75 & 0.99 & 65.90 & 4.05 & 1.01 & 86.25
& 2.95 & 0.88 \\
0.17$\mathbbm{1}_{4,3}$ & 82.75 & 3.00 & 1.59 & 75.85 & 3.05 & 1.62 & 86.90
& 2.70 & 1.46 \\
0.17$\mathbbm{1}_{4,4}$ & 77.40 & 0 & 2.27 & 78.50 & 0 & 2.29 & 66.80 & 0 &
2.04 \\ \hline
& \multicolumn{9}{c}{$\rho_{ij}=0.9, i\neq j$} \\ \hline
$\mathbbm{1}_{4,0}$ & 4.50 & 3.40 & 0 & 4.80 & 4.80 & 0 & 4.20 & 1.40 & 0 \\
0.11$\mathbbm{1}_{4,1}$ & 90.65 & 3.65 & 0.21 & 26.65 & 4.25 & 0.25 & 95.05
& 2.40 & 0.33 \\
0.09$\mathbbm{1}_{4,2}$ & 86.70 & 4.85 & 0.31 & 23.05 & 5.15 & 0.35 & 92.20
& 3.95 & 0.32 \\
0.1$\mathbbm{1}_{4,3}$ & 84.50 & 3.50 & 0.59 & 29.90 & 3.90 & 0.65 & 89.90 &
5.35 & 0.41 \\
0.2$\mathbbm{1}_{4,4}$ & 81.70 & 0 & 3.02 & 83.25 & 0 & 3.05 & 63.15 & 0 &
2.29 \\ \hline
& \multicolumn{9}{c}{$\rho_{ij}=0.5^{\left\vert i-j\right\vert }$} \\ \hline
$\mathbbm{1}_{4,0}$ & 3.90 & 3.75 & 0 & 4.30 & 4.30 & 0 & 3.70 & 1.45 & 0 \\
0.24$\mathbbm{1}_{4,1}$ & 87.60 & 4.85 & 0.81 & 83.30 & 4.95 & 0.82 & 89.50
& 2.55 & 0.76 \\
0.2$\mathbbm{1}_{4,2}$ & 84.10 & 4.50 & 1.34 & 80.50 & 4.55 & 1.35 & 84.40 &
3.40 & 1.18 \\
0.18$\mathbbm{1}_{4,3}$ & 81.05 & 3.60 & 1.68 & 79.30 & 3.65 & 1.70 & 81.15
& 3.95 & 1.49 \\
0.17$\mathbbm{1}_{4,4}$ & 82.00 & 0 & 2.22 & 83.00 & 0 & 2.24 & 78.30 & 0 &
2.15 \\ \hline
& \multicolumn{9}{c}{$\rho_{ij}=(-0.5)^{\left\vert i-j\right\vert }$} \\
\hline
$\mathbbm{1}_{4,0}$ & 4.95 & 4.45 & 0 & 5.15 & 5.15 & 0 & 4.95 & 3.30 & 0 \\
0.24$\mathbbm{1}_{4,1}$ & 86.05 & 5.00 & 0.80 & 82.95 & 5.30 & 0.82 & 88.25
& 7.85 & 0.79 \\
0.13$\mathbbm{1}_{4,2}$ & 76.20 & 3.45 & 0.50 & 49.95 & 3.60 & 0.53 & 85.25
& 6.15 & 0.54 \\
0.11$\mathbbm{1}_{4,3}$ & 86.40 & 1.95 & 0.55 & 49.05 & 2.00 & 0.56 & 94.15
& 3.00 & 0.70 \\
0.09$\mathbbm{1}_{4,4}$ & 81.10 & 0 & 0.48 & 41.00 & 0 & 0.50 & 90.55 & 0 &
0.66 \\ \hline
\end{tabular}%
\end{center}
\end{table*}

\begin{table*}[tbp]
\caption{The estimated sizes and powers (in percentages) in global testing, and
the estimated FWER (in percentages) and the ANCR (average number of correctly rejected false $H_{i}$) in multiple
testing procedures based on bootstrap with $n=100$.}
\label{tb5}
\begin{center}
\tabcolsep=0.11cm
\par
\begin{tabular}{lccccccccc}
\hline
\multicolumn{1}{c}{$\mu$} & \multicolumn{3}{c}{Combined} &
\multicolumn{3}{c}{MinP} & \multicolumn{3}{c}{$T^{2}$} \\ \hline
& \multicolumn{1}{c}{$H_{K}$} & \multicolumn{1}{c}{FWER} &
\multicolumn{1}{c}{ANCR} & \multicolumn{1}{c}{$H_{K}$} & \multicolumn{1}{c}{
FWER} & \multicolumn{1}{c}{ANCR} & \multicolumn{1}{c}{$H_{K}$} &
\multicolumn{1}{c}{FWER} & \multicolumn{1}{c}{ANCR} \\ \hline
& \multicolumn{9}{c}{$\rho_{ij}=-0.15, i\neq j$} \\ \hline
$\mathbbm{1}_{6,0}$ & 5.00 & 4.85 & 0 & 5.05 & 5.05 & 0 & 5.15 & 1.70 & 0 \\
0.22$\mathbbm{1}_{6,1}$ & 40.35 & 5.20 & 0.34 & 37.70 & 5.50 & 0.34 & 42.10
& 2.30 & 0.19 \\
0.13$\mathbbm{1}_{6,3}$ & 50.20 & 2.85 & 0.29 & 29.25 & 2.90 & 0.30 & 72.30
& 1.85 & 0.19 \\
0.1$\mathbbm{1}_{6,5}$ & 69.05 & 1.40 & 0.25 & 24.95 & 1.45 & 0.26 & 87.55 &
0.85 & 0.18 \\
0.07$\mathbbm{1}_{6,6}$ & 43.55 & 0 & 0.18 & 17.25 & 0 & 0.18 & 67.40 & 0 &
0.11 \\ \hline
& \multicolumn{9}{c}{$\rho_{ij}=0, i\neq j$} \\ \hline
$\mathbbm{1}_{6,0}$ & 5.55 & 5.30 & 0 & 5.50 & 5.50 & 0 & 5.00 & 1.75 & 0 \\
0.25$\mathbbm{1}_{6,1}$ & 46.65 & 4.10 & 0.44 & 46.25 & 4.10 & 0.45 & 37.50
& 1.60 & 0.27 \\
0.2$\mathbbm{1}_{6,3}$ & 63.95 & 3.75 & 0.81 & 61.45 & 3.80 & 0.82 & 70.60 &
3.00 & 0.60 \\
0.15$\mathbbm{1}_{6,5}$ & 54.05 & 0.75 & 0.67 & 48.85 & 0.75 & 0.68 & 65.70
& 1.20 & 0.61 \\
0.14$\mathbbm{1}_{6,6}$ & 54.30 & 0 & 0.68 & 49.40 & 0 & 0.69 & 68.15 & 0 &
0.65 \\ \hline
& \multicolumn{9}{c}{$\rho_{ij}=0.5, i\neq j$} \\ \hline
$\mathbbm{1}_{6,0}$ & 4.90 & 4.70 & 0 & 5.25 & 5.25 & 0 & 4.05 & 0.90 & 0 \\
0.22$\mathbbm{1}_{6,1}$ & 43.35 & 5.50 & 0.34 & 37.25 & 5.60 & 0.35 & 50.20
& 1.80 & 0.27 \\
0.17$\mathbbm{1}_{6,3}$ & 48.80 & 2.60 & 0.58 & 39.55 & 2.85 & 0.59 & 59.95
& 2.20 & 0.40 \\
0.17$\mathbbm{1}_{6,5}$ & 49.45 & 0.60 & 0.89 & 47.80 & 0.60 & 0.90 & 49.55
& 1.80 & 0.60 \\
0.17$\mathbbm{1}_{6,6}$ & 49.85 & 0 & 1.05 & 51.25 & 0 & 1.06 & 29.90 & 0 &
0.76 \\ \hline
& \multicolumn{9}{c}{$\rho_{ij}=0.9, i\neq j$} \\ \hline
$\mathbbm{1}_{6,0}$ & 6.10 & 5.10 & 0 & 6.05 & 6.05 & 0 & 4.75 & 1.00 & 0 \\
0.11$\mathbbm{1}_{6,1}$ & 44.60 & 4.75 & 0.10 & 13.90 & 5.35 & 0.11 & 59.60
& 0.60 & 0.13 \\
0.09$\mathbbm{1}_{6,3}$ & 54.50 & 1.40 & 0.19 & 12.75 & 2.10 & 0.22 & 69.65
& 1.05 & 0.11 \\
0.1$\mathbbm{1}_{6,5}$ & 41.85 & 1.00 & 0.35 & 17.75 & 1.20 & 0.38 & 56.45 &
3.60 & 0.16 \\
0.2$\mathbbm{1}_{6,6}$ & 49.15 & 0 & 1.32 & 52.10 & 0 & 1.37 & 25.60 & 0 &
0.87 \\
& \multicolumn{9}{c}{$\rho_{ij}=0.5^{\left\vert i-j\right\vert }$} \\ \hline
$\mathbbm{1}_{6,0}$ & 6.35 & 6.15 & 0 & 6.30 & 6.30 & 0 & 5.40 & 2.15 & 0 \\
0.24$\mathbbm{1}_{6,1}$ & 46.70 & 4.65 & 0.41 & 44.65 & 4.90 & 0.42 & 46.60
& 1.90 & 0.26 \\
0.2$\mathbbm{1}_{6,3}$ & 53.15 & 2.85 & 0.84 & 53.25 & 2.90 & 0.85 & 49.20 &
1.85 & 0.53 \\
0.18$\mathbbm{1}_{6,5}$ & 57.80 & 0.80 & 1.03 & 57.20 & 0.80 & 1.04 & 52.70
& 2.05 & 0.80 \\
0.17$\mathbbm{1}_{6,6}$ & 55.05 & 0 & 1.06 & 55.80 & 0 & 1.07 & 46.20 & 0 &
0.91 \\ \hline
& \multicolumn{9}{c}{$\rho_{ij}=(-0.5)^{\left\vert i-j\right\vert }$} \\
\hline
$\mathbbm{1}_{6,0}$ & 5.55 & 5.25 & 0 & 5.60 & 5.60 & 0 & 4.85 & 1.35 & 0 \\
0.24$\mathbbm{1}_{6,1}$ & 44.70 & 4.50 & 0.40 & 42.25 & 4.60 & 0.40 & 47.40
& 2.15 & 0.26 \\
0.13$\mathbbm{1}_{6,3}$ & 49.90 & 3.05 & 0.27 & 28.05 & 3.15 & 0.28 & 71.55
& 1.95 & 0.18 \\
0.11$\mathbbm{1}_{6,5}$ & 65.30 & 1.00 & 0.34 & 30.60 & 1.00 & 0.35 & 85.15
& 0.55 & 0.27 \\
0.09$\mathbbm{1}_{6,6}$ & 51.70 & 0 & 0.26 & 22.90 & 0 & 0.27 & 72.70 & 0 &
0.19 \\ \hline
\end{tabular}%
\end{center}
\end{table*}

\begin{table*}[tbp]
\caption{The estimated sizes and powers (in percentages) in global testing, and
the estimated FWER (in percentages) and the ANCR (average number of correctly rejected false $H_{i}$) in multiple
testing procedures based on bootstrap with $n=200$.}
\label{tb6}
\begin{center}
\tabcolsep=0.11cm
\par
\begin{tabular}{lccccccccc}
\hline
\multicolumn{1}{c}{$\mu$} & \multicolumn{3}{c}{Combined} &
\multicolumn{3}{c}{MinP} & \multicolumn{3}{c}{$T^{2}$} \\ \hline
& \multicolumn{1}{c}{$H_{K}$} & \multicolumn{1}{c}{FWER} &
\multicolumn{1}{c}{ANCR} & \multicolumn{1}{c}{$H_{K}$} & \multicolumn{1}{c}{
FWER} & \multicolumn{1}{c}{ANCR} & \multicolumn{1}{c}{$H_{K}$} &
\multicolumn{1}{c}{FWER} & \multicolumn{1}{c}{ANCR} \\ \hline
& \multicolumn{9}{c}{$\rho_{ij}=-0.15, i\neq j$} \\ \hline
$\mathbbm{1}_{6,0}$ & 5.85 & 5.75 & 0 & 6.10 & 6.10 & 0 & 5.90 & 1.95 & 0 \\
0.22$\mathbbm{1}_{6,1}$ & 74.20 & 4.35 & 0.68 & 69.80 & 4.40 & 0.69 & 78.05
& 2.25 & 0.55 \\
0.13$\mathbbm{1}_{6,3}$ & 89.70 & 3.05 & 0.65 & 54.20 & 3.10 & 0.66 & 96.70
& 2.75 & 0.51 \\
0.1$\mathbbm{1}_{6,5}$ & 98.40 & 0.95 & 0.57 & 48.95 & 0.95 & 0.58 & 99.90 &
1.05 & 0.60 \\
0.07$\mathbbm{1}_{6,6}$ & 88.20 & 0 & 0.33 & 30.40 & 0 & 0.34 & 96.65 & 0 &
0.28 \\ \hline
& \multicolumn{9}{c}{$\rho_{ij}=0, i\neq j$} \\ \hline
$\mathbbm{1}_{6,0}$ & 4.85 & 4.75 & 0 & 4.80 & 4.80 & 0 & 4.30 & 1.50 & 0 \\
0.25$\mathbbm{1}_{6,1}$ & 83.20 & 4.95 & 0.83 & 83.30 & 4.95 & 0.83 & 74.40
& 2.40 & 0.67 \\
0.2$\mathbbm{1}_{6,3}$ & 93.35 & 2.95 & 1.81 & 92.00 & 2.95 & 1.81 & 96.45 &
3.95 & 1.55 \\
0.15$\mathbbm{1}_{6,5}$ & 88.80 & 1.00 & 1.59 & 82.75 & 1.00 & 1.60 & 94.70
& 3.55 & 1.79 \\
0.14$\mathbbm{1}_{6,6}$ & 90.60 & 0 & 1.62 & 82.95 & 0 & 1.62 & 96.50 & 0 &
2.17 \\ \hline
& \multicolumn{9}{c}{$\rho_{ij}=0.5, i\neq j$} \\ \hline
$\mathbbm{1}_{6,0}$ & 4.65 & 4.30 & 0 & 4.60 & 4.60 & 0 & 4.45 & 1.10 & 0 \\
0.22$\mathbbm{1}_{6,1}$ & 82.30 & 5.15 & 0.71 & 73.60 & 5.35 & 0.71 & 87.20
& 2.20 & 0.74 \\
0.17$\mathbbm{1}_{6,3}$ & 86.10 & 2.20 & 1.32 & 67.65 & 2.35 & 1.33 & 92.60
& 3.30 & 1.20 \\
0.17$\mathbbm{1}_{6,5}$ & 82.30 & 0.30 & 1.83 & 77.35 & 0.35 & 1.84 & 87.00
& 4.40 & 2.01 \\
0.17$\mathbbm{1}_{6,6}$ & 79.10 & 0 & 1.98 & 79.70 & 0 & 1.99 & 61.45 & 0 &
2.30 \\ \hline
& \multicolumn{9}{c}{$\rho_{ij}=0.9, i\neq j$} \\ \hline
$\mathbbm{1}_{6,0}$ & 4.85 & 3.85 & 0 & 5.05 & 5.05 & 0 & 4.50 & 0.75 & 0 \\
0.11$\mathbbm{1}_{6,1}$ & 86.20 & 4.80 & 0.22 & 26.25 & 5.15 & 0.24 & 93.40
& 1.55 & 0.33 \\
0.09$\mathbbm{1}_{6,3}$ & 93.60 & 2.25 & 0.44 & 24.20 & 2.50 & 0.48 & 97.30
& 1.90 & 0.30 \\
0.1$\mathbbm{1}_{6,5}$ & 80.75 & 1.15 & 0.68 & 30.90 & 1.45 & 0.73 & 89.90 &
6.00 & 0.39 \\
0.2$\mathbbm{1}_{6,6}$ & 81.20 & 0 & 2.34 & 83.50 & 0 & 2.38 & 56.60 & 0 &
2.62 \\ \hline
& \multicolumn{9}{c}{$\rho_{ij}=0.5^{\left\vert i-j\right\vert }$} \\ \hline
$\mathbbm{1}_{6,0}$ & 5.60 & 5.35 & 0 & 5.60 & 5.60 & 0 & 4.90 & 1.70 & 0 \\
0.24$\mathbbm{1}_{6,1}$ & 82.75 & 4.75 & 0.79 & 80.20 & 4.75 & 0.79 & 84.05
& 1.85 & 0.69 \\
0.2$\mathbbm{1}_{6,3}$ & 86.20 & 2.60 & 1.82 & 84.90 & 2.65 & 1.83 & 85.55 &
2.70 & 1.53 \\
0.18$\mathbbm{1}_{6,5}$ & 88.80 & 0.75 & 2.11 & 88.15 & 0.75 & 2.11 & 88.90
& 3.90 & 2.45 \\
0.17$\mathbbm{1}_{6,6}$ & 87.30 & 0 & 2.15 & 87.55 & 0 & 2.16 & 82.95 & 0 &
2.89 \\ \hline
& \multicolumn{9}{c}{$\rho_{ij}=(-0.5)^{\left\vert i-j\right\vert }$} \\
\hline
$\mathbbm{1}_{6,0}$ & 5.40 & 5.25 & 0 & 5.60 & 5.60 & 0 & 6.20 & 2.10 & 0 \\
0.24$\mathbbm{1}_{6,1}$ & 84.20 & 4.80 & 0.79 & 80.80 & 4.90 & 0.80 & 84.40
& 1.90 & 0.68 \\
0.13$\mathbbm{1}_{6,3}$ & 91.95 & 3.20 & 0.68 & 57.10 & 3.30 & 0.69 & 97.35
& 2.30 & 0.54 \\
0.11$\mathbbm{1}_{6,5}$ & 97.80 & 0.70 & 0.77 & 58.90 & 0.70 & 0.79 & 99.40
& 1.05 & 0.81 \\
0.09$\mathbbm{1}_{6,6}$ & 92.85 & 0 & 0.58 & 46.00 & 0 & 0.60 & 98.20 & 0 &
0.60 \\ \hline
\end{tabular}%
\end{center}
\end{table*}

\begin{table*}[tbp]
\caption{The estimated sizes and powers (in percentages) in global testing, and
the estimated FWER (in percentages) and the ANCR (average number of correctly rejected false $H_{i}$) in multiple
testing procedures based on the limiting normal distribution with $k=10$ and $n=100$.}
\label{tb7}
\begin{center}
\tabcolsep=0.11cm
\par
\begin{tabular}{lccccccc}
\hline
\multicolumn{1}{c}{$\mu$} & \multicolumn{3}{c}{Combined} &
\multicolumn{3}{c}{MinP} & \multicolumn{1}{c}{$T^{2}$} \\ \hline
& \multicolumn{1}{c}{$H_{K}$} & \multicolumn{1}{c}{FWER} &
\multicolumn{1}{c}{ANCR} & \multicolumn{1}{c}{$H_{K}$} & \multicolumn{1}{c}{
FWER} & \multicolumn{1}{c}{ANCR} & \multicolumn{1}{c}{$H_{K}$} \\ \hline
& \multicolumn{7}{c}{$\rho_{ij}=-0.1, i\neq j$} \\ \hline
$\mathbbm{1}_{10,0}$ 	&	6.00	&	5.90	&	0	&	5.90	&	5.90	&	0	&	4.90	\\
0.17$\mathbbm{1}_{10,1}$ 	&	17.80	&	4.80	&	0.12	&	16.90	&	4.90	&	0.13	&	22.10	\\
0.1$\mathbbm{1}_{10,3}$ 	&	26.20	&	4.30	&	0.11	&	14.70	&	4.40	&	0.11	&	52.60	\\
0.07$\mathbbm{1}_{10,5}$ 	&	29.60	&	2.30	&	0.09	&	10.60	&	2.30	&	0.09	&	59.20	\\
0.05$\mathbbm{1}_{10,7}$ 	&	28.00	&	1.80	&	0.09	&	10.10	&	1.80	&	0.09	&	59.80	\\
0.04$\mathbbm{1}_{10,10}$ 	&	40.10	&	0	&	0.09	&	9.30	&	0	&	0.10	&	73.60	\\
\hline
& \multicolumn{7}{c}{$\rho_{ij}=0, i\neq j$} \\ \hline
$\mathbbm{1}_{10,0}$ 	&	5.20	&	5.10	&	0	&	5.30	&	5.30	&	0	&	4.90	\\
0.2$\mathbbm{1}_{10,1}$ 	&	22.50	&	5.40	&	0.18	&	22.30	&	5.40	&	0.18	&	16.60	\\
0.15$\mathbbm{1}_{10,3}$ 	&	30.50	&	4.00	&	0.30	&	29.70	&	4.00	&	0.30	&	32.70	\\
0.13$\mathbbm{1}_{10,5}$ 	&	30.30	&	1.80	&	0.32	&	28.60	&	2.10	&	0.32	&	39.20	\\
0.15$\mathbbm{1}_{10,7}$ 	&	55.00	&	1.40	&	0.67	&	50.70	&	1.60	&	0.68	&	70.90	\\
0.1$\mathbbm{1}_{10,10}$ 	&	31.90	&	0	&	0.35	&	30.00	&	0	&	0.36	&	47.50	\\
\hline
& \multicolumn{7}{c}{$\rho_{ij}=0.5, i\neq j$} \\ \hline
$\mathbbm{1}_{10,0}$ 	&	5.30	&	5.00	&	0	&	5.30	&	5.30	&	0	&	4.70	\\
0.19$\mathbbm{1}_{10,1}$ 	&	24.60	&	4.80	&	0.20	&	22.80	&	5.00	&	0.20	&	29.70	\\
0.15$\mathbbm{1}_{10,3}$ 	&	30.60	&	4.90	&	0.31	&	24.80	&	5.20	&	0.32	&	44.70	\\
0.14$\mathbbm{1}_{10,5}$ 	&	37.90	&	4.00	&	0.52	&	31.30	&	4.20	&	0.53	&	53.20	\\
0.13$\mathbbm{1}_{10,7}$ 	&	31.80	&	1.30	&	0.53	&	28.00	&	1.50	&	0.54	&	39.60	\\
0.13$\mathbbm{1}_{10,10}$ 	&	31.70	&	0	&	0.81	&	32.90	&	0	&	0.82	&	13.90	\\
 \hline
& \multicolumn{7}{c}{$\rho_{ij}=0.9, i\neq j$} \\ \hline
$\mathbbm{1}_{10,0}$ 	&	4.50	&	3.50	&	0	&	4.70	&	4.70	&	0	&	4.90	\\
0.09$\mathbbm{1}_{10,1}$ 	&	20.00	&	5.90	&	0.06	&	10.50	&	6.60	&	0.07	&	32.90	\\
0.06$\mathbbm{1}_{10,3}$ 	&	23.00	&	4.60	&	0.11	&	9.70	&	5.60	&	0.13	&	36.80	\\
0.06$\mathbbm{1}_{10,5}$ 	&	25.80	&	2.90	&	0.15	&	7.90	&	3.00	&	0.16	&	40.90	\\
0.07$\mathbbm{1}_{10,7}$ 	&	29.20	&	1.80	&	0.31	&	9.40	&	1.90	&	0.32	&	48.00	\\
0.2$\mathbbm{1}_{10,10}$ 	&	49.40	&	0	&	2.83	&	51.70	&	0	&	2.87	&	19.50	\\
 \hline
& \multicolumn{7}{c}{$\rho_{ij}=0.5^{\left\vert i-j\right\vert }$} \\ \hline
$\mathbbm{1}_{10,0}$ 	&	5.50	&	5.40	&	0	&	5.40	&	5.40	&	0	&	4.90	\\
0.2$\mathbbm{1}_{10,1}$ 	&	22.90	&	5.80	&	0.18	&	22.50	&	5.90	&	0.18	&	22.30	\\
0.2$\mathbbm{1}_{10,3}$ 	&	45.90	&	3.90	&	0.68	&	45.80	&	4.00	&	0.68	&	39.10	\\
0.2$\mathbbm{1}_{10,5}$ 	&	56.90	&	3.20	&	1.10	&	57.00	&	3.60	&	1.11	&	50.00	\\
0.15$\mathbbm{1}_{10,7}$ 	&	41.60	&	1.80	&	0.74	&	42.00	&	1.80	&	0.75	&	33.90	\\
0.1$\mathbbm{1}_{10,10}$ 	&	26.10	&	0	&	0.38	&	26.70	&	0	&	0.38	&	17.40	\\
 \hline
& \multicolumn{7}{c}{$\rho_{ij}=(-0.5)^{\left\vert i-j\right\vert }$} \\
\hline
$\mathbbm{1}_{10,0}$ 	&	4.60	&	4.30	&	0	&	4.50	&	4.50	&	0	&	4.90	\\
0.18$\mathbbm{1}_{10,1}$ 	&	19.30	&	5.20	&	0.14	&	19.20	&	5.40	&	0.14	&	18.00	\\
0.11$\mathbbm{1}_{10,3}$ 	&	24.80	&	4.00	&	0.14	&	17.80	&	4.50	&	0.15	&	43.10	\\
0.08$\mathbbm{1}_{10,5}$ 	&	19.30	&	2.90	&	0.10	&	12.30	&	3.00	&	0.11	&	39.30	\\
0.08$\mathbbm{1}_{10,7}$ 	&	29.80	&	1.60	&	0.17	&	16.80	&	1.70	&	0.17	&	56.00	\\
0.07$\mathbbm{1}_{10,10}$ 	&	35.70	&	0	&	0.20	&	17.80	&	0	&	0.20	&	63.20	\\
 \hline
\end{tabular}%
\end{center}
\end{table*}

\begin{table*}[tbp]
\caption{The estimated sizes and powers (in percentages) in global testing, and
the estimated FWER (in percentages) and the ANCR (average number of correctly rejected false $H_{i}$) in multiple
testing procedures based on the limiting normal distribution with $k=10$ and $n=200$.}
\label{tb8}
\begin{center}
\tabcolsep=0.11cm
\par
\begin{tabular}{lccccccc}
\hline
\multicolumn{1}{c}{$\mu$} & \multicolumn{3}{c}{Combined} &
\multicolumn{3}{c}{MinP} & \multicolumn{1}{c}{$T^{2}$} \\ \hline
& \multicolumn{1}{c}{$H_{K}$} & \multicolumn{1}{c}{FWER} &
\multicolumn{1}{c}{ANCR} & \multicolumn{1}{c}{$H_{K}$} & \multicolumn{1}{c}{
FWER} & \multicolumn{1}{c}{ANCR} & \multicolumn{1}{c}{$H_{K}$} \\ \hline
& \multicolumn{7}{c}{$\rho_{ij}=-0.1, i\neq j$} \\ \hline
$\mathbbm{1}_{10,0}$ 	&	5.50	&	5.30	&	0	&	5.40	&	5.40	&	0	&	5.30	\\
0.17$\mathbbm{1}_{10,1}$ 	&	42.50	&	4.60	&	0.35	&	38.00	&	4.80	&	0.35	&	53.20	\\
0.1$\mathbbm{1}_{10,3}$ 	&	70.00	&	4.00	&	0.26	&	27.40	&	4.00	&	0.26	&	91.10	\\
0.07$\mathbbm{1}_{10,5}$ 	&	80.00	&	2.20	&	0.18	&	18.40	&	2.20	&	0.18	&	94.70	\\
0.05$\mathbbm{1}_{10,7}$ 	&	77.70	&	1.60	&	0.13	&	14.70	&	1.60	&	0.14	&	93.40	\\
0.04$\mathbbm{1}_{10,10}$ 	&	90.40	&	0	&	0.12	&	12.00	&	0	&	0.13	&	98.60	\\
\hline
& \multicolumn{7}{c}{$\rho_{ij}=0, i\neq j$} \\ \hline
$\mathbbm{1}_{10,0}$ 	&	5.90	&	5.70	&	0	&	5.80	&	5.80	&	0	&	5.30	\\
0.2$\mathbbm{1}_{10,1}$ 	&	53.70	&	3.30	&	0.52	&	53.50	&	3.60	&	0.52	&	40.40	\\
0.15$\mathbbm{1}_{10,3}$ 	&	61.20	&	3.10	&	0.78	&	60.20	&	3.10	&	0.78	&	69.50	\\
0.13$\mathbbm{1}_{10,5}$ 	&	66.20	&	3.80	&	0.89	&	61.00	&	3.90	&	0.89	&	78.40	\\
0.15$\mathbbm{1}_{10,7}$ 	&	93.90	&	2.40	&	1.87	&	86.40	&	2.40	&	1.87	&	98.60	\\
0.1$\mathbbm{1}_{10,10}$ 	&	69.90	&	0	&	0.84	&	57.80	&	0	&	0.85	&	86.90	\\
\hline
& \multicolumn{7}{c}{$\rho_{ij}=0.5, i\neq j$} \\ \hline
$\mathbbm{1}_{10,0}$ 	&	4.90	&	4.50	&	0	&	4.70	&	4.70	&	0	&	5.30	\\
0.19$\mathbbm{1}_{10,1}$ 	&	56.50	&	4.70	&	0.48	&	51.60	&	4.70	&	0.50	&	65.70	\\
0.15$\mathbbm{1}_{10,3}$ 	&	71.90	&	3.30	&	0.89	&	51.90	&	3.30	&	0.90	&	86.20	\\
0.14$\mathbbm{1}_{10,5}$ 	&	76.90	&	3.00	&	1.24	&	56.30	&	3.00	&	1.25	&	89.50	\\
0.13$\mathbbm{1}_{10,7}$ 	&	67.10	&	2.90	&	1.50	&	54.50	&	2.90	&	1.51	&	78.70	\\
0.13$\mathbbm{1}_{10,10}$ 	&	59.90	&	0	&	2.06	&	60.50	&	0	&	2.06	&	32.40	\\
 \hline
& \multicolumn{7}{c}{$\rho_{ij}=0.9, i\neq j$} \\ \hline
$\mathbbm{1}_{10,0}$ 	&	5.80	&	4.30	&	0.00	&	5.30	&	5.30	&	0.00	&	5.30	\\
0.09$\mathbbm{1}_{10,1}$ 	&	53.40	&	3.80	&	0.11	&	14.70	&	4.00	&	0.13	&	69.70	\\
0.06$\mathbbm{1}_{10,3}$ 	&	56.80	&	3.20	&	0.16	&	12.30	&	3.80	&	0.18	&	74.80	\\
0.06$\mathbbm{1}_{10,5}$ 	&	68.70	&	3.80	&	0.35	&	13.60	&	3.80	&	0.37	&	83.20	\\
0.07$\mathbbm{1}_{10,7}$ 	&	78.50	&	1.70	&	0.65	&	18.60	&	1.90	&	0.66	&	87.90	\\
0.2$\mathbbm{1}_{10,10}$ 	&	78.10	&	0.00	&	5.06	&	80.70	&	0.00	&	5.09	&	45.60	\\
 \hline
& \multicolumn{7}{c}{$\rho_{ij}=0.5^{\left\vert i-j\right\vert }$} \\ \hline
$\mathbbm{1}_{10,0}$ 	&	4.50	&	4.40	&	0	&	4.70	&	4.70	&	0	&	5.30	\\
0.2$\mathbbm{1}_{10,1}$ 	&	57.30	&	5.40	&	0.52	&	55.50	&	5.50	&	0.53	&	52.80	\\
0.2$\mathbbm{1}_{10,3}$ 	&	80.10	&	3.90	&	1.61	&	79.50	&	3.90	&	1.61	&	77.60	\\
0.2$\mathbbm{1}_{10,5}$ 	&	91.00	&	4.30	&	2.71	&	90.30	&	4.30	&	2.71	&	87.80	\\
0.15$\mathbbm{1}_{10,7}$ 	&	75.40	&	2.10	&	1.90	&	75.00	&	2.10	&	1.91	&	73.50	\\
0.1$\mathbbm{1}_{10,10}$ 	&	47.60	&	0	&	0.95	&	47.70	&	0	&	0.95	&	39.50	\\
 \hline
& \multicolumn{7}{c}{$\rho_{ij}=(-0.5)^{\left\vert i-j\right\vert }$} \\
\hline
$\mathbbm{1}_{10,0}$ 	&	6.00	&	5.90	&	0	&	6.30	&	6.30	&	0	&	5.30	\\
0.18$\mathbbm{1}_{10,1}$ 	&	45.90	&	4.50	&	0.41	&	44.60	&	4.60	&	0.41	&	44.30	\\
0.11$\mathbbm{1}_{10,3}$ 	&	60.40	&	3.70	&	0.33	&	32.90	&	3.80	&	0.33	&	81.90	\\
0.08$\mathbbm{1}_{10,5}$ 	&	55.20	&	2.90	&	0.25	&	24.50	&	3.10	&	0.26	&	80.10	\\
0.08$\mathbbm{1}_{10,7}$ 	&	78.80	&	1.90	&	0.36	&	31.20	&	1.90	&	0.36	&	93.80	\\
0.07$\mathbbm{1}_{10,10}$ 	&	83.40	&	0	&	0.37	&	31.00	&	0	&	0.37	&	96.00	\\
 \hline
\end{tabular}%
\end{center}
\end{table*}

\begin{table*}[tbp]
\caption{The estimated sizes and powers (in percentages) in global testing, and
the estimated FWER (in percentages) and the ANCR (average number of correctly rejected false $H_{i}$) in multiple
testing procedures based on the limiting normal distribution with $k=20$.}
\label{tb9}
\begin{center}
\tabcolsep=0.11cm
\par
\begin{tabular}{lccccccc}
\hline
\multicolumn{1}{c}{$\mu$} & \multicolumn{3}{c}{Combined} &
\multicolumn{3}{c}{MinP} & \multicolumn{1}{c}{$T^{2}$} \\ \hline
& \multicolumn{1}{c}{$H_{K}$} & \multicolumn{1}{c}{FWER} &
\multicolumn{1}{c}{ANCR} & \multicolumn{1}{c}{$H_{K}$} & \multicolumn{1}{c}{
FWER} & \multicolumn{1}{c}{ANCR} & \multicolumn{1}{c}{$H_{K}$} \\ \hline
& \multicolumn{7}{c}{$\rho_{ij}=0, i\neq j$, $n=200$} \\ \hline
$\mathbbm{1}_{20,0}$ 	&	5.90	&	5.20	&	0	&	5.40	&	5.40	&	0	&	5.70	\\
0.18$\mathbbm{1}_{20,1}$ 	&	35.20	&	3.90	&	0.31	&	34.60	&	3.90	&	0.32	&	21.30	\\
0.12$\mathbbm{1}_{20,1}$ 	&	47.90	&	3.50	&	0.48	&	41.80	&	3.80	&	0.49	&	55.40	\\
0.08$\mathbbm{1}_{20,1}$ 	&	41.30	&	3.50	&	0.32	&	30.50	&	3.80	&	0.32	&	48.40	\\
0.07$\mathbbm{1}_{20,1}$ 	&	44.60	&	1.80	&	0.33	&	29.20	&	1.80	&	0.33	&	55.50	\\
0.07$\mathbbm{1}_{20,1}$ 	&	59.10	&	0	&	0.44	&	36.50	&	0	&	0.45	&	71.00	\\
\hline
& \multicolumn{7}{c}{$\rho_{ij}=0.5, i\neq j$, $n=400$} \\ \hline
$\mathbbm{1}_{20,0}$ 	&	5.80	&	5.30	&	0	&	5.50	&	5.50	&	0	&	5.30	\\
0.3$\mathbbm{1}_{20,1}$ 	&	51.40	&	5.70	&	0.41	&	44.20	&	5.70	&	0.42	&	56.30	\\
0.1$\mathbbm{1}_{20,1}$ 	&	68.10	&	4.60	&	0.64	&	38.40	&	4.60	&	0.65	&	82.60	\\
0.1$\mathbbm{1}_{20,1}$ 	&	54.10	&	3.10	&	0.62	&	29.10	&	3.10	&	0.62	&	72.00	\\
0.06$\mathbbm{1}_{20,1}$ 	&	65.30	&	1.90	&	1.47	&	46.40	&	2.00	&	1.48	&	78.10	\\
0.08$\mathbbm{1}_{20,1}$ 	&	48.30	&	0	&	2.19	&	48.50	&	0	&	2.20	&	17.90	\\
 \hline
& \multicolumn{7}{c}{$\rho_{ij}=0.9, i\neq j$, $n=400$} \\ \hline
$\mathbbm{1}_{20,0}$ 	&	6.30	&	4.80	&	0	&	5.30	&	5.30	&	0	&	4.90	\\
0.05$\mathbbm{1}_{20,1}$ 	&	43.10	&	3.40	&	0.58	&	11.60	&	3.50	&	0.60	&	53.80	\\
0.03$\mathbbm{1}_{20,1}$ 	&	44.30	&	5.10	&	0.17	&	9.40	&	5.20	&	0.18	&	56.00	\\
0.03$\mathbbm{1}_{20,1}$ 	&	56.90	&	3.10	&	0.27	&	7.90	&	3.50	&	0.28	&	67.20	\\
0.03$\mathbbm{1}_{20,1}$ 	&	43.10	&	3.40	&	0.58	&	11.60	&	3.50	&	0.60	&	53.80	\\
0.08$\mathbbm{1}_{20,1}$ 	&	38.10	&	0	&	4.34	&	39.60	&	0	&	4.37	&	12.20	\\
 \hline
& \multicolumn{7}{c}{$\rho_{ij}=0.5^{\left\vert i-j\right\vert }$, $n=200$} \\ \hline
$\mathbbm{1}_{20,0}$ 	&	6.20	&	5.50	&	0	&	5.40	&	5.40	&	0	&	5.40	\\
0.18$\mathbbm{1}_{20,1}$ 	&	40.10	&	4.40	&	0.34	&	36.40	&	4.30	&	0.34	&	28.90	\\
0.13$\mathbbm{1}_{20,1}$ 	&	47.20	&	5.30	&	0.65	&	44.20	&	5.30	&	0.65	&	33.50	\\
0.1$\mathbbm{1}_{20,1}$ 	&	41.80	&	2.30	&	0.63	&	40.00	&	2.30	&	0.63	&	32.60	\\
0.08$\mathbbm{1}_{20,1}$ 	&	35.70	&	1.90	&	0.50	&	33.10	&	1.90	&	0.51	&	28.20	\\
0.07$\mathbbm{1}_{20,1}$ 	&	33.60	&	0	&	0.49	&	32.80	&	0	&	0.49	&	23.50	\\
 \hline
& \multicolumn{7}{c}{$\rho_{ij}=(-0.5)^{\left\vert i-j\right\vert }$, $n=200$} \\
\hline
$\mathbbm{1}_{20,0}$ 	&	6.70	&	6.00	&	0	&	6.00	&	6.00	&	0	&	5.80	\\
0.2$\mathbbm{1}_{20,1}$ 	&	52.80	&	6.90	&	0.46	&	48.80	&	6.60	&	0.46	&	42.20	\\
0.08$\mathbbm{1}_{20,1}$ 	&	46.90	&	4.30	&	0.16	&	17.90	&	4.30	&	0.16	&	64.50	\\
0.05$\mathbbm{1}_{20,1}$ 	&	37.60	&	2.50	&	0.13	&	13.70	&	2.30	&	0.13	&	52.50	\\
0.04$\mathbbm{1}_{20,1}$ 	&	36.40	&	1.50	&	0.12	&	12.50	&	1.50	&	0.12	&	51.90	\\
0.04$\mathbbm{1}_{20,1}$ 	&	53.10	&	0	&	0.19	&	17.10	&	0	&	0.19	&	67.00	\\
 \hline
\end{tabular}%
\end{center}
\end{table*}

\begin{table*}[tbp]
\caption{The estimated sizes and powers (in percentages) in global testing, and
the estimated FWER (in percentages) and the ANCR (average number of correctly rejected false $H_{i}$) in multiple
testing procedures based on the limiting normal distribution with $k=20$.}
\label{tb10}
\begin{center}
\tabcolsep=0.11cm
\par
\begin{tabular}{lccccccc}
\hline
\multicolumn{1}{c}{$\mu$} & \multicolumn{3}{c}{Combined} &
\multicolumn{3}{c}{MinP} & \multicolumn{1}{c}{$T^{2}$} \\ \hline
& \multicolumn{1}{c}{$H_{K}$} & \multicolumn{1}{c}{FWER} &
\multicolumn{1}{c}{ANCR} & \multicolumn{1}{c}{$H_{K}$} & \multicolumn{1}{c}{
FWER} & \multicolumn{1}{c}{ANCR} & \multicolumn{1}{c}{$H_{K}$} \\ \hline
& \multicolumn{7}{c}{$\rho_{ij}=0, i\neq j$, $n=400$} \\ \hline
$\mathbbm{1}_{20,0}$ 	&	6.60	&	6.00	&	0	&	6.20	&	6.20	&	0	&	6.20	\\
0.18$\mathbbm{1}_{20,1}$ 	&	73.90	&	5.20	&	0.72	&	74.40	&	5.20	&	0.73	&	49.70	\\
0.12$\mathbbm{1}_{20,1}$ 	&	85.10	&	4.40	&	1.37	&	78.90	&	4.50	&	1.37	&	91.50	\\
0.08$\mathbbm{1}_{20,1}$ 	&	74.80	&	4.00	&	0.83	&	58.40	&	4.00	&	0.83	&	88.30	\\
0.07$\mathbbm{1}_{20,1}$ 	&	79.00	&	1.90	&	0.86	&	57.90	&	1.90	&	0.86	&	91.30	\\
0.07$\mathbbm{1}_{20,1}$ 	&	92.60	&	0	&	1.15	&	69.20	&	0	&	1.15	&	98.10	\\
\hline
& \multicolumn{7}{c}{$\rho_{ij}=0.5, i\neq j$, $n=500$} \\ \hline
$\mathbbm{1}_{20,0}$ 	&	5.80	&	5.30	&	0	&	5.50	&	5.50	&	0	&	5.30	\\
0.3$\mathbbm{1}_{20,1}$ 	&	51.40	&	5.70	&	0.41	&	44.20	&	5.70	&	0.42	&	56.30	\\
0.1$\mathbbm{1}_{20,1}$ 	&	68.10	&	4.60	&	0.64	&	38.40	&	4.60	&	0.65	&	82.60	\\
0.1$\mathbbm{1}_{20,1}$ 	&	54.10	&	3.10	&	0.62	&	29.10	&	3.10	&	0.62	&	72.00	\\
0.06$\mathbbm{1}_{20,1}$ 	&	65.30	&	1.90	&	1.47	&	46.40	&	2.00	&	1.48	&	78.10	\\
0.08$\mathbbm{1}_{20,1}$ 	&	48.30	&	0	&	2.19	&	48.50	&	0	&	2.20	&	17.90	\\
 \hline
& \multicolumn{7}{c}{$\rho_{ij}=0.9, i\neq j$, $n=500$} \\ \hline
$\mathbbm{1}_{20,0}$ 	&	6.00	&	4.30	&	0	&	4.90	&	4.90	&	0	&	3.50	\\
0.05$\mathbbm{1}_{20,1}$ 	&	34.30	&	4.90	&	0.07	&	11.10	&	5.40	&	0.08	&	46.30	\\
0.03$\mathbbm{1}_{20,1}$ 	&	20.20	&	4.20	&	0.12	&	6.00	&	4.60	&	0.13	&	29.70	\\
0.03$\mathbbm{1}_{20,1}$ 	&	71.50	&	4.30	&	0.38	&	10.10	&	4.30	&	0.39	&	81.40	\\
0.03$\mathbbm{1}_{20,1}$ 	&	53.90	&	2.70	&	0.51	&	9.70	&	2.80	&	0.52	&	67.70	\\
0.08$\mathbbm{1}_{20,1}$ 	&	44.90	&	0	&	6.00	&	46.00	&	0	&	6.04	&	15.40	\\
 \hline
& \multicolumn{7}{c}{$\rho_{ij}=0.5^{\left\vert i-j\right\vert }$, $n=400$} \\ \hline
$\mathbbm{1}_{20,0}$ 	&	5.90	&	5.30	&	0	&	5.20	&	5.20	&	0	&	5.20	\\
0.18$\mathbbm{1}_{20,1}$ 	&	74.10	&	4.80	&	0.72	&	73.00	&	4.90	&	0.72	&	64.60	\\
0.13$\mathbbm{1}_{20,1}$ 	&	77.70	&	5.30	&	1.81	&	77.30	&	5.20	&	1.82	&	69.40	\\
0.1$\mathbbm{1}_{20,1}$ 	&	71.70	&	3.30	&	1.69	&	70.70	&	3.40	&	1.69	&	65.70	\\
0.08$\mathbbm{1}_{20,1}$ 	&	63.30	&	2.10	&	1.28	&	61.60	&	2.10	&	1.28	&	61.30	\\
0.07$\mathbbm{1}_{20,1}$ 	&	60.80	&	0	&	1.22	&	59.00	&	0	&	1.22	&	57.30	\\
 \hline
& \multicolumn{7}{c}{$\rho_{ij}=(-0.5)^{\left\vert i-j\right\vert }$, $n=400$} \\
\hline
$\mathbbm{1}_{20,0}$ 	&	6.00	&	5.20	&	0	&	5.00	&	5.00	&	0	&	5.40	\\
0.2$\mathbbm{1}_{20,1}$ 	&	86.10	&	4.30	&	0.83	&	84.30	&	4.30	&	0.83	&	78.30	\\
0.08$\mathbbm{1}_{20,1}$ 	&	88.10	&	4.60	&	0.44	&	40.20	&	4.50	&	0.44	&	96.90	\\
0.05$\mathbbm{1}_{20,1}$ 	&	75.40	&	2.30	&	0.21	&	20.90	&	2.30	&	0.21	&	92.20	\\
0.04$\mathbbm{1}_{20,1}$ 	&	74.40	&	1.90	&	0.23	&	21.10	&	1.90	&	0.23	&	92.50	\\
0.04$\mathbbm{1}_{20,1}$ 	&	91.90	&	0	&	0.30	&	26.30	&	0	&	0.30	&	98.80	\\
 \hline
\end{tabular}%
\end{center}
\end{table*}

\end{document}